\def\appendix{\par
 \setcounter{section}{0}
 \setcounter{subsection}{0}
 \def\thesection{Appendix \Alph{section}}
 \def\thesubsection{\Alph{section}.\arabic{subsection}}
 \def\theequation{\Alph{section}.\arabic{equation}}
 \setcounter{equation}{0}}

\documentstyle[preprint,aps,floats,epsfig]{revtex}
\begin{document}


\title{Deep inelastic Scattering, QCD, and generalised vector
  dominance\footnote{Supported by the Bundesministerium f\"ur Bildung und
  Forschung, Bonn, Germany.}}

\author{G.~Cveti\v{c}$^{a,b}$\footnote{
cvetic@physik.uni-bielefeld.de},
D.~Schildknecht$^{a}$\footnote{
Dieter.Schildknecht@physik.uni-bielefeld.de} 
and A.~Shoshi$^{a}$\footnote{
shoshi@physik.uni-bielefeld.de}
}

\address{ $^a$Dept.~of Physics, Bielefeld University,
              33501 Bielefeld, Germany}
\address{ $^b$Dept.~of Physics, Dortmund University,
              44221 Dortmund, Germany}

\begin{flushright}
BI-TP 99/28 \\
hep-ph/9908473
\end{flushright}

\vspace{1.5cm}

\centerline{{\large \bf
Deep inelastic Scattering, QCD, and generalised vector dominance}}

\vspace{0.8cm}

\centerline{ G.~Cveti\v{c}, D.~Schildknecht and A.~Shoshi}

\newpage

\maketitle

\renewcommand{\thefootnote}{\fnsymbol{footnote}}
\renewcommand{\thefootnote}{\arabic{footnote}}
\begin{abstract}
We provide a formulation of generalised vector dominance (GVD) for low--$x$
deep-inelastic scattering that explicitly incorporates the ${\gamma}^{\ast} \to
q{\bar q}$ transition and a QCD-inspired ansatz for the $(q{\bar q})p$
forward-scattering amplitude. The destructive interference originally 
introduced in off-diagonal GVD is recovered in the present formulation and
traced back to the generic structure of two-gluon-exchange
as incorporated into the notion of colour transparency.
Asymptotically, the transverse photoabsorption cross section behaves as
$(\ln Q^2)/Q^2$, implying a logarithmic violation of scaling for $F_2$, while
the longitudinal-to-transverse ratio decreases as $1/\ln Q^2$.
\end{abstract}
\newpage
\section{Introduction}

The observation of diffractive production of high-mass states at HERA
\cite{HERA} 
at small values of the scaling variable $x \approx {Q}^2/{W}^2$
qualitatively confirms the expectation from generalised vector 
dominance (GVD) \cite{SakuraiSchildknecht}\footnote{Compare also 
\cite{Gribov} for a formulation of GVD for complex nuclei.}. 
The starting point of GVD is provided by a mass
dispersion relation, the spectral weight function therein containing the
coupling of a vector
state of mass $M_V$ to a timelike photon, as observed in 
${e}^{+}\!{e}^{-}$ annihilation, and the forward scattering 
of the vector state from the nucleon. 

Originating from the pre-QCD era, the coupling of the photon to the
high-mass continuum of ${e}^{+}\!{e}^{-}$ annihilation is
frequently described in a global effective manner; the $q{\bar q}$ jets
originating from the ${\gamma}^{\ast}\!\to\!q{\bar q}$ coupling, as observed at
sufficiently high energies in ${e}^{+}\!{e}^{-}$ annihilation, are not
explicitly incorporated into the description \cite{SchildknechtSpiesberger} of deep inelastic scattering.

In the present work, we provide a formulation of GVD that 
quantitatively takes into account not only the energy dependence of the ${\gamma}^{\ast}{\to}q{\bar q}$
transition, but the dependence on the $q{\bar q}$ configuration as well within the spectral weight function of GVD. The ansatz for 
the subsequent scattering of the $q{\bar q}$ state will be inspired by
QCD. The emphasis of the present work will be put on the general theoretical
analysis. Even though numerical results will be given, it will not be the aim of
the present work to carry out a detailed comparison with the experimental data.

In Sec.~\ref{section2}, we formulate the virtual Compton forward amplitude
in terms of the  ${\gamma}^{\ast} \to q{\bar q}$ transition of a timelike
photon, continued to the spacelike region via appropriate propagator
factors, and an ansatz motivated by perturbative QCD (pQCD)
for the $(q{\bar q})p$ forward scattering
amplitude. The destructive interference originally
incorporated into off-diagonal GVD \cite{FRS} reappears as the essential 
feature of the pQCD-inspired ansatz.

In Sec.~\ref{section3}, the results of Sec.~\ref{section2} are rederived in
transverse position space, using the notion of colour transparency.

In Sections \ref{section4} and \ref{section5}, we explicitly present the
consequences from the QCD-inspired GVD ansatz for the $Q^2$ dependence of the
transverse and the longitudinal photon-absorption cross section. 

Some conclusions are drawn in section \ref{section6}.

\section{Off-diagonal Generalised Vector Dominance from QCD.}
\label{section2}

The GVD picture for the Compton forward amplitude is described in Fig.1.
We start with the ${\gamma}^{\ast}{\to}q{\bar q}$ 
transition. We look at the transition of a timelike photon of mass
${q}^{2}{\equiv}{M}_{q{\bar q}}^{2}$ to the $q{\bar q}$ pair. 
\begin{figure}[htb]
\setlength{\unitlength}{1.cm}
\begin{center}
\epsfig{file=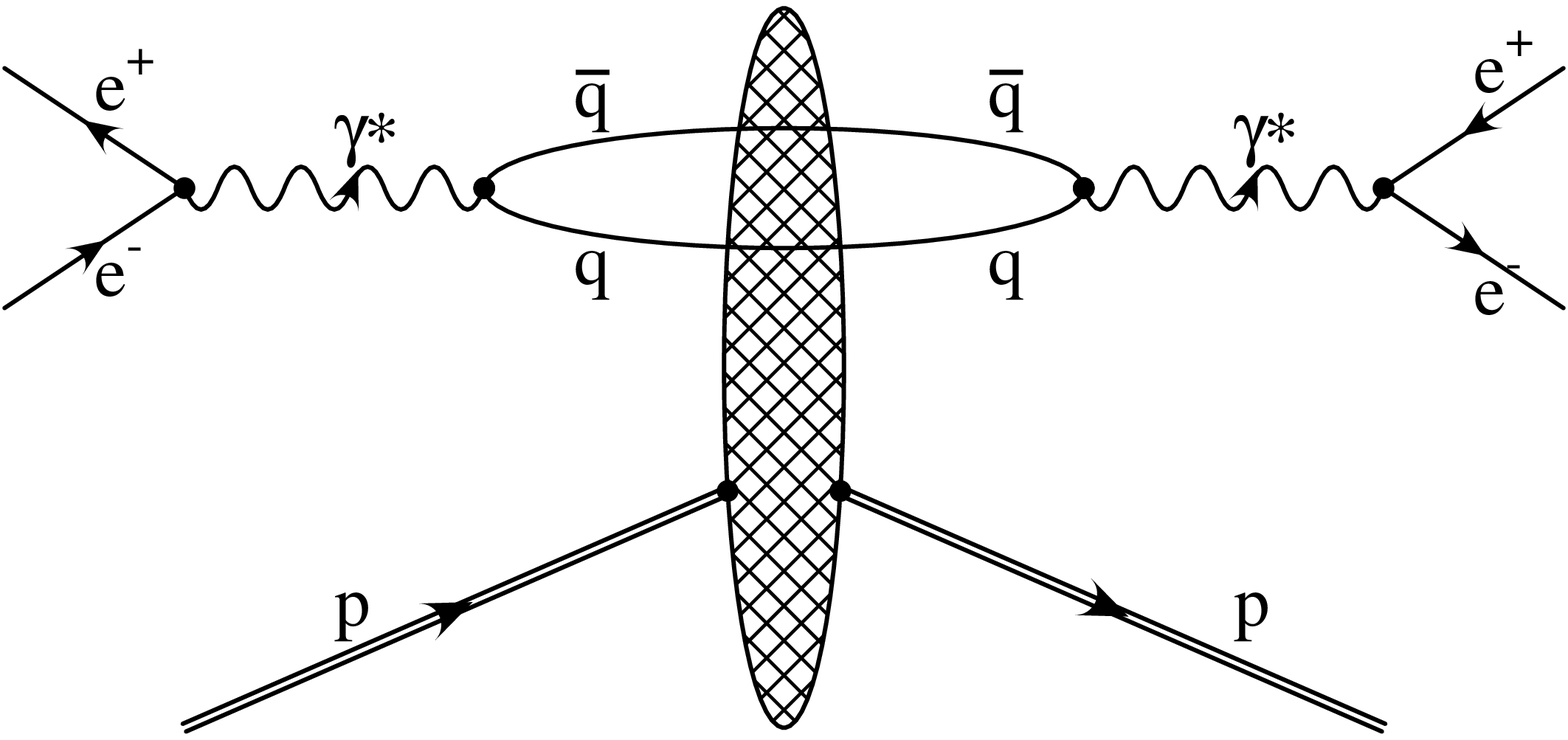,width=10.cm}
 \begin{center} a) \end{center}
\epsfig{file=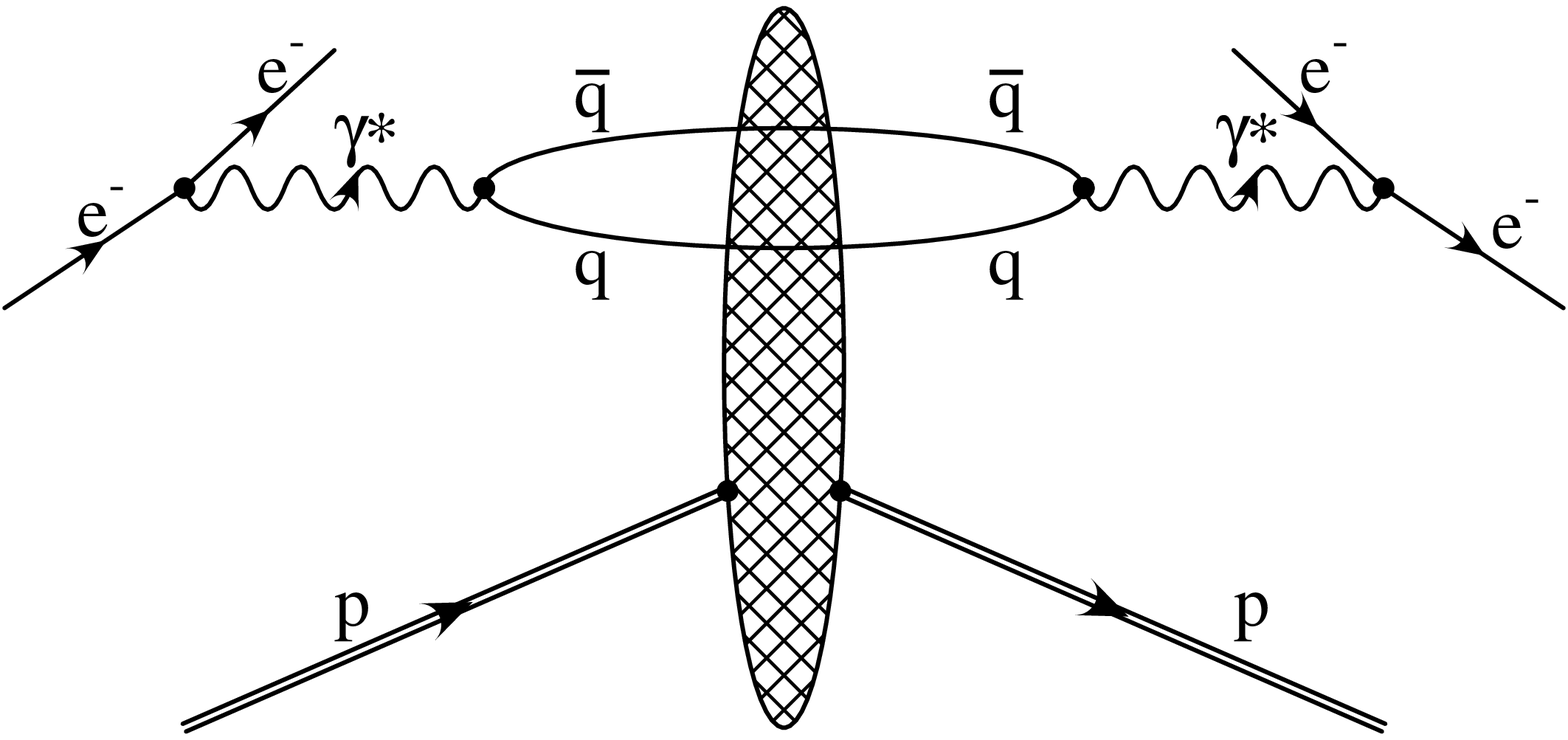,width=10.cm}
 \begin{center} b) \end{center}
\end{center}
\caption{\small The Compton forward amplitude in the proton rest frame, 
a) in the Gedankenexperiment where a timelike photon of mass 
$q^2=M_{q{\bar q}}^2$ interacts with the nucleon, 
b) upon continuation from $q^2=M_{q{\bar q}}^2$ to
$q^2=-Q^2<0$, with $x(\approx Q^2/W^2){\ll}~1$.}
\label{Fig1}
\end{figure}
The four-momentum
of the photon of mass ${q}^{2}{\equiv}{M}_{q{\bar q}}^{2}$ in its rest frame is
given by ${q}^{\mu}={k}^{\mu}_{q}+{k}^{\mu}_{\bar q}=({M}_{q{\bar q}},{\vec
  0})$, where ${k}^{\mu}_{q}$ and ${k}^{\mu}_{\bar q}$ denote the four-momenta
of the quark and antiquark, respectively.

The $q{\bar q}$ current may be written as\footnote{Here, we work in the 
approximation of massless quarks.} 
\begin{equation}
{\bar {\it u}}^{(\lambda)}({k}_{q}){\gamma}^{\mu}{\it v}^{({\lambda}^{\prime})}({k}_{\bar
  q})=-{M}_{q{\bar q}}(0,\cos\vartheta\cos\varphi+{\rm
  i}\lambda\sin\varphi,\cos\vartheta\sin\varphi-{\rm
  i}\lambda\cos\varphi,-\sin\vartheta){\delta}_{\lambda,-{\lambda}^{\prime}} \ .
\end{equation}
Here, $\vartheta$ and $\varphi$ denote the polar and azimuthal production 
angles of the quark with respect to the $z$--axis in the photon rest frame, 
${\vec k}_q=|{\vec k}_q|\left (\sin\vartheta \cos\varphi, \sin\vartheta \sin\varphi,
\cos\vartheta \right)$, ${\vec k}_{\perp}=|{\vec k}_{\perp}|\left(
\cos\varphi, \sin\varphi \right)$, and
$\lambda$, ${\lambda}^{\prime}$ denote twice the quark and antiquark
helicities. The timelike photon is supposed to originate from the annihilation
of an ${e}^{+}{e}^{-}$ pair, and the $z$--axis is chosen in the direction 
of the ${e}^{-}$ three-momentum in the 
${e}^{+}{e}^{-}$ (photon) rest frame (cf. Fig.2.). 
\begin{figure}[htb]
\setlength{\unitlength}{1.cm}
\begin{center}
\epsfig{file=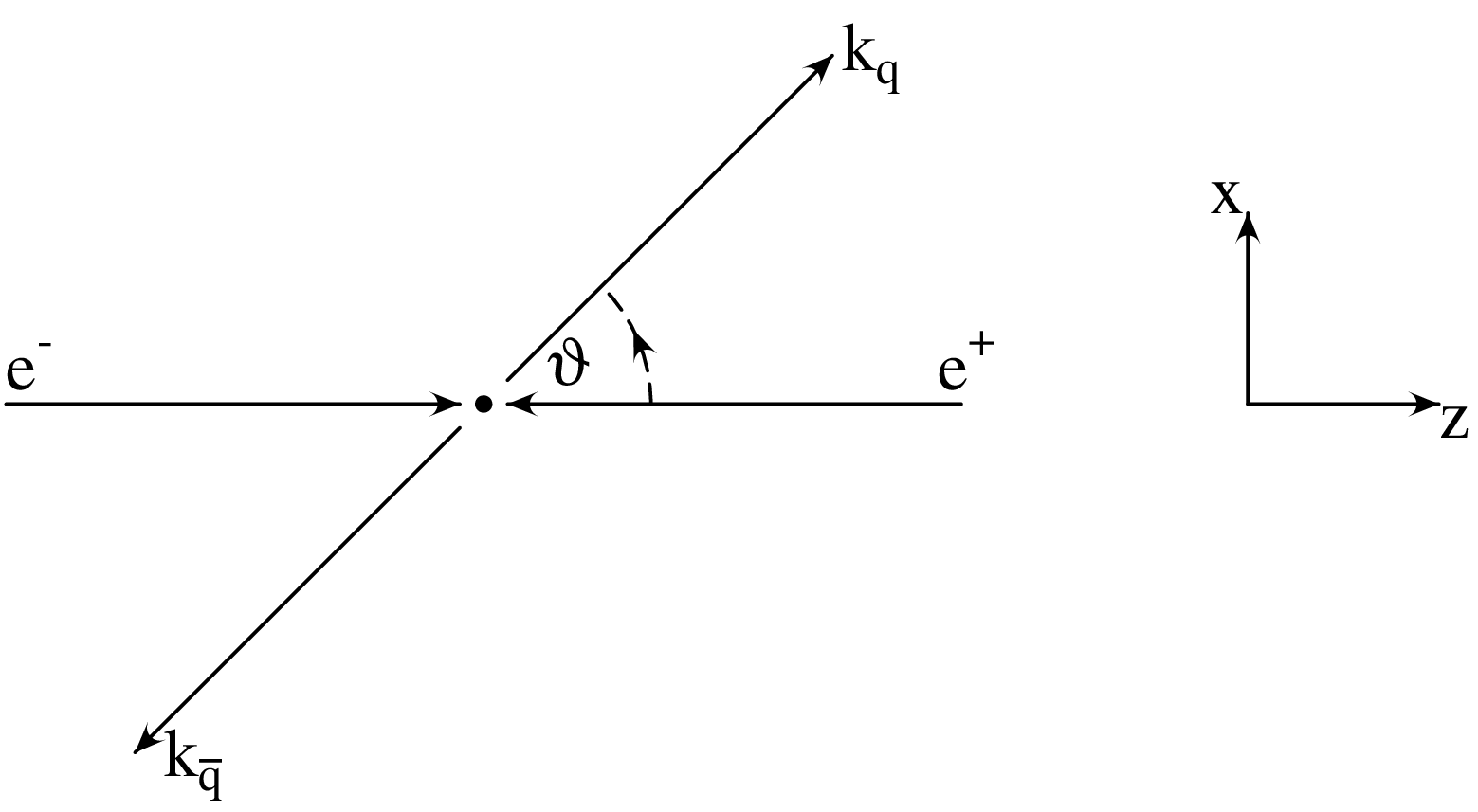,width=10cm}
\end{center}
\caption{\small The process of ${e}^{+}{e}^{-}$ annihilation in the rest frame
  of the $q{\bar q}$ system.}
\label{Fig2}
\end{figure}
The assumed origin of the timelike photon from ${e}^{+}{e}^{-}$ annihilation
(obviously) not only defines the four-momentum, but the
polarisation properties of the photon as well. Introducing longitudinal and transverse
helicity states for the massive photon in its rest frame,
\begin{equation}
{\epsilon}_{L}^{\mu}=(0,0,0,1), \quad
{\epsilon}_{T}^{\mu}(\pm)=\frac{1}{\sqrt{2}}(0,1,\pm{\rm i},0),
\label{epsilon}
\end{equation}
one obtains
\begin{eqnarray}
{\it j}_{L} &\equiv & {\bar {\it
  u}}^{(\lambda)}({k}_{q}){\epsilon}_{{\mu}L}{\gamma}^{\mu}{\it
  v}^{({\lambda}^{\prime})}({k}_{\bar q})=-{M}_{q{\bar
  q}}\sin\vartheta{\delta}_{{\lambda},-{{\lambda}^{\prime}}} \ ,
\nonumber\\
{\it j}_{T}(\pm) &\equiv & {\bar {\it
  u}}^{(\lambda)}({k}_{q}){\epsilon}_{{\mu}T}(\pm){\gamma}^{\mu}{\it
  v}^{({\lambda}^{\prime})}({k}_{\bar q})=\frac{{M}_{q{\bar
  q}}}{\sqrt{2}}e^{\pm{\rm i}\varphi}
(\cos\vartheta\pm\lambda){\delta}_{\lambda,-{\lambda}^{\prime}}.
\label{j}
\end{eqnarray}
Substituting
\begin{eqnarray}
{\sin}^{2}\vartheta &\equiv& \frac{4{m}_{\perp}^{2}}{{M}^{2}_{q{\bar
      q}}}=4z(1-z)\ ,\quad \cos\vartheta=(2z-1) \ ,
\label{jj}
\end{eqnarray}
where
\begin{eqnarray}
z=\frac{1}{2} \pm \frac{1}{2}\sqrt{1-\frac{4{m}^{2}_{\perp}}{{M}^{2}_{q{\bar
        q}}}},\quad m_{\perp}^2 \equiv k^2_{\perp}, \quad(0~{\leq}~z~{\leq}~1) \ ,
\label{z}
\end{eqnarray}
one may represent (\ref{j}) in a manifestly covariant form
\begin{eqnarray}
{\it j}_{L} &=& -{M}_{q{\bar
    q}}2\sqrt{z(1\!-\!z)}{\delta}_{{\lambda},-{\lambda}^{\prime}} \ ,
\nonumber\\
{\it j}_{T}(\pm) &=& \frac{{M}_{q{\bar q}}}{\sqrt{2}}e^{\pm{\it
  i}\varphi}(2z-1\pm\lambda){\delta}_{{\lambda},-{\lambda}^{\prime}} \ .
\label{j1}
\end{eqnarray}
Longitudinal and transverse components of the current are thus explicitly
defined with respect to any Lorentz frame obtained from the rest frame by a Lorentz
boost in the $z$ direction. In particular, when considering the forward amplitude
for scattering of the $q{\bar q}$ state from the nucleon at high energies, an
appropriate Lorentz boost in the $z$ direction is to be applied to the $q{\bar q}$
system. Incidentally, we note that $z$ from (\ref{z}) can also be represented as
\begin{equation}
z = \frac{k^0 + k^3}{q^0+q^3}, \quad (k^{\mu} \equiv k_q^{\mu}) \ .
\label{zfrac}
\end{equation}
Hence, $z$ is unchanged under Lorentz boosts along the
photon direction. In the high-energy limit, $|q^0| \approx |q^3| \gg M_{q{\bar
    q}}$, $z$ becomes identical to the fraction of the 
longitudinal momentum \cite{wavef} of the $q{\bar q}$ system carried by
the quark $q$.

Since
\begin{equation}
{M}^{2}_{q{\bar q}}=\frac{{k}_{\perp}^{2}}{z(1\!-\!z)} \ ,
\label{M}
\end{equation}
we can, instead of the pair of variables $({M}_{q{\bar q}}, z)$ characterising the
$q{\bar q}$ state coupled to the timelike photon, alternatively use 
$(k^2_{\perp}, z)$ in (\ref{j1}).

Relations (\ref{j1}), upon multiplication by the antiquark charge $(-e_q)$, give the 
coupling to the (timelike) photon of the $q{\bar q}$ complex of mass
${M}_{q{\bar q}}$ (or, alternatively, the transverse momentum 
${\vec k}_{\perp}$) and the additional ``configuration'' 
degree of freedom, $z$. Let us envisage a
physical situation in which such a high-energy $q{\bar q}$ complex, 
originating from a timelike photon, hits the proton in its rest frame
(Fig.1a). Continuing\footnote{
Compare e.g. Ref.~\cite{Brodsky} for a detailed discussion
on the lifetime arguments \cite{Ioffe} relevant in connection with the
continuation to spacelike $q^2$.} 
to spacelike four-momenta of the photon,
$q^2{\equiv}-Q^2<0$, with\footnote{
Here, $W^2=(q+p)^2$, where $p$ is the four-momentum of the proton.}
$x \approx Q^2/W^2 \ll 1$,
requires multiplication of the $q{\bar q}$ forward scattering amplitude by the
coupling to the photon from (\ref{j1}) and a propagator
factor $1/(Q^2+M^2_{q{\bar q}})$. 

At this point, the cases of transverse and longitudinal photons have to be
discriminated. For transverse photons, one simply assumes that the 
dependence on $Q^2$  induced by the propagator is the only one in the 
high-energy limit with $x \approx Q^2/W^2 \ll 1$. Accordingly, 
\begin{equation}
{\it A}_{{\gamma}^{\ast}p{\to}{\gamma}^{\ast}p} \sim
{\overline {\cal M}}^{\ast}_{T} {\cal T}_{(q{\bar q})p \to (q{\bar q})p}
{\overline {\cal M}}_{T} \ ,
 \label{Aq}
\end{equation}
where the $(q{\bar q})p$ forward scattering amplitude is denoted by 
${\cal T}_{(q{\bar q})p \to (q{\bar q})p}$, 
and according to (\ref{j1}) and the above discussion
\begin{equation}
{\overline {\cal M}}^{({\lambda},{\lambda}^{\prime})}_{T}
({M}_{q{\bar q}},z,Q^2)=
- \frac{{e}_{q}}{Q^2+{M}_{q{\bar q}}^{2}}\frac{{M}_{q{\bar
    q}}}{\sqrt{2}}e^{\pm{\it
    i}\varphi}(2z-1\pm{\lambda}){\delta}_{\lambda,-{\lambda}^{\prime}} \ .
\label{j2}
\end{equation}

For longitudinal photons, the restriction that the photon couples to a 
conserved source leads to a $Q^2$ dependence in addition to the one induced by
the propagator. Current conservation requires that the $q{\bar q}$ system
couples to a conserved source. This leads \cite{FS} to an additional factor
$ \sqrt{ Q^2/M^2_{q{\bar q}} }$. Even though this factor is related to
the $(q{\bar q})p$ amplitude and not to the ${\gamma}^{\ast}~\to~q{\bar q}$
transition, it may be put together with the propagator to yield 
\begin{equation}
{\overline {\cal M}}^{({\lambda},{\lambda}^{\prime})}_{L}(M_{q{\bar q}},z,Q^2)=
-\frac{e_q}{Q^2+M_{q{\bar q}}^2} M_{q{\bar q}}
\sqrt{\frac{Q^2}{M^2_{q{\bar q}}}} 2 
\sqrt{z(1-z)}{\delta}_{\lambda,-{\lambda}^{\prime}}.
\label{j3}
\end{equation}

Inclusion of a quark mass, $m_q$, changes (\ref{M}) to become 
$M^2_{q{\bar q}} = (k_{\perp}^2\!+\!m_q^2)/(z(1-z))$. The transverse transition
amplitude (\ref{j2}) is modified by an additive term proportional to $m_q$,
\begin{equation}
{\overline {\cal M}}^{({\lambda},{\lambda}^{\prime})}_{T}({M}_{q{\bar q}},z,Q^2)=
- \frac{{e}_{q}}{(Q^2+M_{q{\bar q}}^2) \sqrt{2}} 
\left[ M_{q{\bar q}} e^{\pm{\rm i}\varphi}
(2z-1\pm{\lambda}){\delta}_{\lambda,-{\lambda}^{\prime}} 
+ \frac{(\lambda \pm 1) m_q}{\sqrt{z(1\!-\!z)}} {\delta}_{\lambda,\lambda^{\prime}}
\right]
\ ,
\label{j2mq}
\end{equation}
while the expression (\ref{j3}) for the longitudinal amplitude remains unchanged.

In terms of the imaginary part of the forward-scattering amplitude\footnote{
A factor $1/W^2$ from the optical theorem is included in 
${\cal T}_{(q{\bar q})p \to (q{\bar q})p}({\vec k}_{\perp}^{\prime}, z^{\prime}; 
{\vec k}_{\perp}, z;W^2)$.} 
${\cal T}$, the total photoabsorption cross section for transverse
(${\gamma}^{\ast}_T$) and longitudinal (${\gamma}_L^{\ast}$) 
virtual photons, via the use of the optical theorem, becomes
\begin{eqnarray}
\lefteqn{
{\sigma}_{{\gamma}^{\ast}_{T,L} p}(W^2,Q^2) =
\left[ \frac{1}{2 (2 \pi)^3} \right]^2
\sum_{\lambda, {\lambda}^{\prime} = \pm 1} 
\int dz \int d{z}^{\prime}\int_{|{\vec k}_{\perp}| 
\ge k_{\perp 0}} d^2 {k}_{\perp} 
\int_{|{\vec k}_{\perp}^{\prime}| \ge k_{\perp 0}} d^2 {k}_{\perp}^{\prime}
\times }
\nonumber\\
&&
{\cal M}^{(\lambda,\lambda^{\prime})}_{T,L}( {\vec k}_{\perp}^{\prime},
z^{\prime};Q^2)^{\ast}
 {\cal T}_{(q{\bar q})p \to (q{\bar q})p}({\vec
  k}_{\perp}^{\prime}, z^{\prime}; {\vec k}_{\perp}, z; W^2) 
{\cal M}^{(\lambda,\lambda^{\prime})}_{T,L}({\vec k}_{\perp}, z;Q^2) \ ,
\label{sigtot}
\end{eqnarray}
where 
\begin{equation}
{\cal M}^{(\lambda,\lambda^{\prime})}_{T,L}({\vec k}_{\perp},z;Q^2) \equiv
\frac{{\overline {\cal M}}^{(\lambda,\lambda^{\prime})}_{T,L}({\vec
  k}_{\perp},z;Q^2)}{\sqrt{z(1-z)}} \ .
\label{Mll}
\end{equation}
The overall factors $1/\sqrt{(z(1\!-\!z))}$  entering (\ref{sigtot})
via (\ref{Mll}) originate from the phase--space 
factors $d^3{k}_{(i)}/( k^0_{(i)} 2 (2 \pi)^3 )$. They were
rewritten using the identities
$d^3{k}_{(i)}/k^0_{(i)} \equiv
d^2{k}_{\perp (i)} dk^3_{(i)}/k^0_{(i)}$
$=d^2{k}_{\perp (i)}  dz_i/z_i$, 
where $z_i$ ($i=1,2$) is the fraction of the longitudinal $q{\bar q}$ 
momentum carried by one of the quarks ($z_1\!=\!z$, $z_2\!=\!1\!-\!z$). 

In (\ref{sigtot}), we have indicated lower limits, $k_{\perp 0}$, for the
integration over the transverse momenta. The lower limit in transverse-momentum  space
corresponds to a finite transverse extension of the $q{\bar q}$ state in 
position space (confinement). The threshold, $k_{\perp 0}$, is introduced, in
order to allow (\ref{sigtot}) to be used in an effective description of
${\sigma}_{{\gamma}^{\ast}_{T,L} p}$ at low values of
$Q^2$, where the low--lying vector mesons actually dominate the
Compton forward amplitude.

\begin{figure*}[htb]
\setlength{\unitlength}{1.cm}
\begin{center}
\epsfig{file=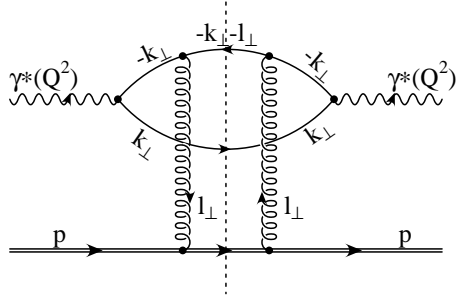,width=7.cm}
 \begin{center} \small a) \end{center}
\vspace{1cm}
\epsfig{file=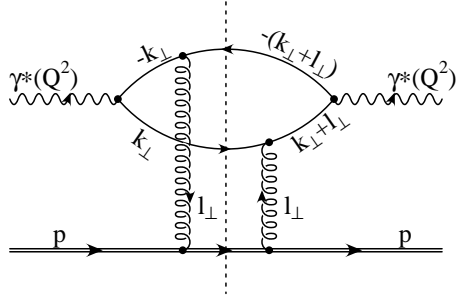,width=7.cm}
 \begin{center} \small b) \end{center}
\caption{\small The two-gluon exchange realisation of the
  structure (\ref{tscat}), (\ref{sigtot1}). The diagrams (a) and (b) correspond
  to transitions diagonal and off-diagonal in the masses of the $q{\bar q}$
  pairs, respectively.} 
\label{Fig3}
\end{center}
\end{figure*}

So far, the $(q{\bar q})p$ scattering amplitude has been left unspecified. To
proceed, we will look for guidance at the two-gluon-exchange \cite{Soper} of
perturbative QCD (pQCD). As illustrated in Fig.~\ref{Fig3}\footnote{Additional
  diagrams are suppressed in Fig.~\ref{Fig3}, as the generic structure of the
  diagrams is our only concern in the present context.}, two-gluon exchange
contains ``diagonal'' as well as ``off-diagonal'' transitions with respect to
the transverse momenta ${\vec k}_{\perp}$ and ${\vec k}_{\perp}+{\vec
  l}_{\perp}$ and the  masses,
\begin{equation}
M^{2}_{q{\bar q}}=\frac{{\vec k}_{\perp}^2}{z(1-z)} \ ,\quad
M^{\prime 2}_{q{\bar q}}=\frac{({\vec k}_{\perp}\!+\!{\vec l}_{\perp})^2}{z(1-z)} \ ,
\label{Mp}
\end{equation}
of the incoming and outgoing $q{\bar q}$ state. Fermion (the quark $q$) and
antifermion (the antiquark ${\bar q}$) couple with opposite sign to the gluon. 
Accordingly, diagonal and off-diagonal transitions contribute with the same
weight, but opposite signs.
Guided by the structure of two-gluon exchange in pQCD, we adopt the following
ansatz for the forward scattering amplitude in (\ref{sigtot})\footnote{
The factor $2 (2 \pi)^3$ appears in (\ref{tscat})  due to the normalisation convention
$\langle k_q | k^{\prime}_q \rangle = 2 (2 \pi)^3 k^0_q 
\delta^{(3)}({\vec k}_q - {\vec k}^{\prime}_q)$ used throughout.}:
\begin{eqnarray}
\lefteqn{
{\cal T}_{(q{\bar q})p \to (q{\bar q})p}({\vec k}_{\perp}^{\prime}, z^{\prime};
{\vec k}_{\perp}, z; W^2) = 
}
\nonumber \\
&& 
2 (2 \pi)^3 \int d^2{l}_{\perp} 
{\tilde \sigma}_{(q{\bar q})p}(l^2_{\perp}, W^2) \left \lbrack \delta(
{\vec k}_{\perp}^{\prime}\!-\!{\vec k}_{\perp})\!-\!\delta({\vec
  k}_{\perp}^{\prime}\!-\!{\vec k}_{\perp}\!-\!{\vec l}_{\perp}) \right \rbrack
\delta(z\!-\!z^{\prime}) \ ,
\label{tscat}
\end{eqnarray}
In addition to the difference in sign between the diagonal and the off-diagonal
term, the ansatz (\ref{tscat}) incorporates low--$x$ (high $W^2$) kinematics;
the scattering is assumed to only affect the transverse momentum, while $z$ remains 
unchanged. Further, the $(q{\bar q})p$ interaction,
${\tilde \sigma}_{(q{\bar q})p}$, is assumed to solely be determined by the
transverse momentum transfer $l^2_{\perp}$ and the c.m.s.--energy $W$.

Substituting (\ref {tscat}) into (\ref{sigtot}) yields
\begin{eqnarray}
{\sigma}_{{\gamma}^{\ast}_{T,L} p}&(&W^2,Q^2) = 
\frac{1}{16 \pi^3}\sum_{\lambda, {\lambda}^{\prime} = \pm 1} 
\int dz \int d{z}^{\prime}\int d^2{l}_{\perp} {\tilde \sigma}_{(q{\bar q})p}(l^2_{\perp}, W^2)
\int_{|{\vec k}_{\perp}| \ge k_{\perp 0}} d^2 {k}_{\perp} 
\int_{|{\vec k}_{\perp}^{\prime}| \ge k_{\perp 0}} d^2 {
  k}_{\perp}^{\prime}
\times
\nonumber\\
&&
{\cal M}^{(\lambda,\lambda^{\prime})}_{T,L}({\vec k}_{\perp}^{\prime},
z^{\prime};Q^2)^{\ast}
{\cal M}^{(\lambda,\lambda^{\prime})}_{T,L}({\vec k}_{\perp}, z;Q^2) 
\left \lbrack \delta(
{\vec k}_{\perp}^{\prime}\!-\!{\vec k}_{\perp})\!-\!\delta({\vec
  k}_{\perp}^{\prime}\!-\!{\vec k}_{\perp}\!-\!{\vec l}_{\perp}) \right \rbrack
\delta(z\!-\!z^{\prime})\ .
\label{sigtot1a}
\end{eqnarray}
Upon integration over $d^2{k}_{\perp}^{\prime}$ and $d{z}^{\prime}$
(\ref{sigtot1a}) becomes\footnote{
In the case of transversely polarised photons, averaging over the
two polarisations $P=\pm 1$ is implicitly understood.}
\begin{eqnarray}
{\sigma}_{{\gamma}^{\ast}_{T,L} p}&(&W^2,Q^2) = 
\frac{1}{16 \pi^3} 
\int dz \int d^2 {l}_{\perp}
{\tilde \sigma}_{(q{\bar q})p}({l}^2_{\perp};W^2) 
{\Bigg \lbrace} \int_{|{\vec k}_{\perp}| \ge k_{\perp 0}} d^2 {k}_{\perp}
\sum_{\lambda, {\lambda}^{\prime} = \pm 1} \left| 
{\cal M}^{(\lambda,{\lambda}^{\prime})}_{T,L} (z,{\vec k}_{\perp}; Q^2) 
\right|^2 
\nonumber\\
&&
-\int_{|{\vec k}_{\perp}| \ge k_{\perp 0}, |{\vec k}_{\perp}\!+\!{\vec l}_{\perp}| \ge
  k_{\perp 0}} d^2 {k}_{\perp}
\sum_{\lambda, {\lambda}^{\prime} = \pm 1}
{\cal M}^{(\lambda,{\lambda}^{\prime})}_{T,L} (z,{\vec k}_{\perp}; Q^2) 
{\cal M}^{(\lambda,{\lambda}^{\prime})}_{T,L} 
(z,{\vec k}_{\perp}\!+\!{\vec l}_{\perp}; Q^2)^{\ast}
{\Bigg \rbrace}  \ .
\label{sigtot1}
\end{eqnarray}
The remarkable difference in sign between the diagonal and the off-diagonal term
in (\ref{sigtot1a}) and (\ref{sigtot1}), abstracted from perturbative QCD, 
actually implies significant cancellations between the contributions 
of the two terms. The first term under the integral in the curly bracket of
(\ref{sigtot1}) is related to the square of the amplitude of the process
$({\gamma}^{\ast} \to q{\bar q})p \to {\rm hadrons}$ for fixed mass $M_{q{\bar
    q}}$ (cf.~(\ref{M})), while the second term, the ``off-diagonal'' one with
the negative sign in front, contains the product of the
  amplitudes for different masses, $M_{q{\bar q}}$ and $M^{\prime}_{q{\bar q}}$ (cf.~(\ref{Mp})).
It is worth noting that a structure of destructive 
interference between contributions diagonal and off-diagonal 
in the $q{\bar q}$ mass, as in (\ref{sigtot1}), was actually suggested 
\cite{FRS} a long time ago, in order to reconcile scaling in $e^+e^-$
annihilation with scaling in the deep inelastic scattering in conjunction with a
reasonable (hadronic) cross section 
for the scattering of $q{\bar q}$--vector--meson 
states on the proton. In the framework of the off-diagonal generalised 
vector dominance model \cite{FRS}, the destructive
interference was associated with the couplings of the photon to massive
$q{\bar q}$--vector--meson states. Within the present pQCD-motivated ansatz
(\ref{tscat}), the destructive interference from off-diagonal GVD is
recovered\footnote{
Compare also refs. \cite{Strikman} and \cite{Shaw}, where
similar conclusions were arrived at.}
and traced back to the opposite
couplings of the gluon to the quark and the antiquark the virtual photon 
has dissociated into.

\section{Position-space formulation, colour transparency.}
\label{section3}

In this Section, we rederive (\ref{sigtot1a}) in a
position-space formulation. As the concept of ``colour transparency'' 
\cite{Soper,NikolaevZakharov} underlying
the position-space formulation may also be motivated by the two-gluon exchange of
perturbative QCD and its generalisation, it will come as no surprise that (\ref{sigtot1a}) will be
recovered from an ansatz in position space.

We start by introducing the transverse position variable ${\vec r}_{\perp}$, conjugate to
${\vec k}_{\perp}$, by forming the Fourier transform of ${\cal
  M}^{(\lambda,{\lambda}^{\prime})}_{T,L}(z,{\vec k}_{\perp};Q^2)$ from (\ref{Mll}),  
\begin{equation}
{\psi}_{T,L}^{(\lambda,{\lambda}^{\prime})}(z,{\vec
  r}_{\perp};Q^2){\equiv}\frac{\sqrt{4\pi}}{16{\pi}^{3}}\int_{k_{\perp 0}}
\!d^2{
  k}_{\perp}\exp{({\rm i}{\vec k}_{\perp}\cdot{\vec r}_{\perp})}{\cal
  M}^{(\lambda,{\lambda}^{\prime})}_{T,L}(z,{\vec k}_{\perp};Q^2) \ .
\label{fj1}
\end{equation}
The function ${\psi}_{T,L}^{(\lambda,{\lambda}^{\prime})}(z,{\vec
  r}_{\perp};Q^2)$ \cite{wavef} has frequently been called  
the ``photon--$q{\bar q}$ wave function'' \cite{NikolaevZakharov}.

The $\delta$--function dependence on the initial and final transverse momenta
${\vec k}_{\perp}$ and ${\vec k}^{\prime}_{\perp}$ in (\ref{sigtot1a}) suggests
to adopt a representation for ${\sigma}_{{\gamma}^{\ast}_{T,L}p}$ in transverse
position space that is diagonal with respect to ${\vec r}_{\perp}$,
\begin{equation}
{\sigma}_{ {\gamma}^{\ast}_{T,L} p }(W^2, Q^2) =
\sum_{\lambda, {\lambda}^{\prime} = \pm 1}
\int dz \int d^2{r}_{\perp} 
\left| \psi_{T,L}^{(\lambda,{\lambda}^{\prime})}(z, {\vec r}_{\perp}; Q^2)
\right|^2 {\sigma}_{(q{\bar q})p}({r}^2_{\perp},W^2)
\ ,
\label{ip1}
\end{equation}
i.e. the cross section ${\sigma}_{{\gamma}^{\ast}_{T,L}p}$ is built up by
multiplying the ``dipole cross section'' \cite{NikolaevZakharov} ${\sigma}_{(q{\bar
    q})p}(r^2_{\perp},W^2)$ by the probability to find the incoming quark and
the incoming antiquark a transverse distance $r_{\perp}$ apart from each other. 
The longitudinal variable $z$ is ``frozen'' during 
the scattering process.

In a further step, we specify the relation between the dipole cross section (in
position space) and the transverse-momentum-transfer function ${\tilde
  \sigma}_{(q{\bar q})p}(l^2_{\perp},W^2)$ in (\ref{sigtot1a}). Requiring the
dipole cross section to vanish for zero separation of quark and antiquark, as
suggested by two-gluon exchange or by colour-neutrality of the $q{\bar q}$
state, we have 
\begin{equation}
{\sigma}_{(q{\bar q})p}({r}_{\perp}^{2},W^2)=\int\!d^2{
  l}_{\perp}{\tilde\sigma}_{(q{\bar q})p}({l}_{\perp}^2,W^2)\left(1-{e}^{-{\rm
  i}{\vec l}_{\perp}\cdot{\vec r}_{\perp}}\right).
\label{fs}
\end{equation}
This ansatz indeed incorporates the required vanishing (colour tansparency
\cite{Soper,NikolaevZakharov}), as $r_{\perp}^2$, for zero separation of quark
and antiquark, $r_{\perp} \to 0$,
\begin{eqnarray}
{\sigma}_{(q{\bar q})p}({r}_{\perp}^2,W^2) & \to & 0 \qquad 
({\rm for} \ {r}_{\perp} \to 0) \ , 
\label{ct}
\end{eqnarray}
as well as a constant limit for $r_{\perp} \to \infty$
\begin{eqnarray}
{\sigma}_{(q{\bar q})p}({r}_{\perp}^2,W^2) & \to & 
{\sigma}^{(\infty)}_{(q{\bar q})p}(W^2) \equiv \
{\sigma}^{(\infty)}_{(q{\bar q})p} \quad
({\rm for} \ r_{\perp} \to \infty) \ ,
\label{cons}
\end{eqnarray}
as the integral over the momentum space function has to be finite. From 
Fourier inversion of (\ref{fs}), 
\begin{equation}
{\tilde\sigma}_{(q{\bar q})p}({l}_{\perp}^2,W^2)=
\frac{1}{{(2\pi)}^{2}}
\int\!d^2{r}_{\perp}{e}^{{\rm i}{\vec l}_{\perp}\cdot{\vec r}_{\perp}}
\left[ {\sigma}_{(q{\bar q})p}^{(\infty)} -
{\sigma}_{(q{\bar q})p}({r}_{\perp}^{2},W^2) \right] 
\label{sun3}
\end{equation}
as well as from (\ref{cons}), we have 
${\tilde \sigma}_{(q{\bar q})p}({l}^2_{\perp}) \to 0$ for $l_{\perp} \to
\infty$. Compare Figs.~4a, b for a sketch of the qualitative behaviour of 
${\sigma}_{(q{\bar q})p}({r}_{\perp}^{2}, W^2)$ for two different simple choices of ${\tilde\sigma}_{(q{\bar q})p}({l}_{\perp}^{2},W^2)$.

Inserting the dipole cross section  (\ref{fs}), the position-space
representation (\ref{ip1}) for
${\sigma}_{{\gamma}^{\ast}_{T,L}p}(r^2_{\perp},W^2)$ becomes
\begin{eqnarray}
\lefteqn{
{\sigma}_{{\gamma}^{\ast}_{T,L} p}(W^2,Q^2) =
}
\nonumber\\
&& \sum_{\lambda, {\lambda}^{\prime} = \pm 1}
\int dz \int d^2 {l}_{\perp}
{\tilde \sigma}_{(q{\bar q})p}({l}^2_{\perp},W^2)
\int d^2{r}_{\perp} 
\left| \psi_{T,L}^{(\lambda,{\lambda}^{\prime})}(z, {\vec r}_{\perp}; Q^2)
\right|^2 \left( 1 - 
e^{ - {\rm i} {\vec l}_{\perp}\!\cdot {\vec r}_{\perp} } \right) \ .
\label{ip2}
\end{eqnarray}
Upon introducing the ${\gamma}^{\ast} \to q{\bar q}$ transition amplitude  
(\ref{fj1}), and integrating over position space, we have 
\begin{eqnarray}
{\sigma}_{{\gamma}^{\ast}_{T,L} p}&(&W^2,Q^2) = 
\frac{1}{16 \pi^3} \sum_{\lambda, {\lambda}^{\prime} = \pm 1}
\int dz \int d^2 {l}_{\perp}
{\tilde \sigma}_{(q{\bar q})p}({l}^2_{\perp},W^2) \int_{|{\vec k}_{\perp}| \geq
  k_{\perp 0}} d^2 {k}_{\perp} \int_{|{\vec k}_{\perp}^{\prime}|
  \geq k_{\perp 0}}d^2 {k}_{\perp}^{\prime}\times
\nonumber\\
& &
{\cal M}_{T,L}^{(\lambda,{\lambda}^{\prime})}(z,{\vec k}_{\perp};Q^2)^{\ast}
{\cal M}_{T,L}^{(\lambda,{\lambda}^{\prime})}(z,{\vec k}_{\perp}^{\prime};Q^2)
\left[ \delta ( {\vec k}_{\perp}^{\prime}\!-\!{\vec k}_{\perp} )
- \delta ( {\vec k}_{\perp}^{\prime}\!-\!{\vec k}_{\perp}\!-\!{\vec l}_
{\perp} ) \right] \ . 
\label{ip4}
\end{eqnarray}
This result for ${\sigma}_{{\gamma}^{\ast}_{T,L} p}$ indeed coincides with expression (\ref{sigtot1a}). 

Similar forms of the dipole cross section 
${\sigma}_{(q{\bar q})p}(r^2_{\perp})$ in (\ref{fs})
are obtained from a $\delta$--function ansatz
and from a Gaussian ansatz for ${\tilde\sigma}_{(q{\bar q})p}({l}_{\perp}^{2})$
[cf.~Figs.~4a, b],
\begin{eqnarray}
{\tilde\sigma}_{(q{\bar q})p}({l}_{\perp}^{2}) &=&
\frac{{\sigma}_{(q{\bar q})p}^{(\infty)}}{\pi}
\delta({l}_{\perp}^{2}-{\Lambda}^2) \quad  \Rightarrow \quad
{\sigma}_{(q{\bar q})p}({r}^{2}_{\perp})={\sigma}_{(q{\bar q})p}^{(\infty)}
\left(1-{J}_{0}({\Lambda}|{\vec r}_{\perp}|)\right) \ ;
\label{delta}
\\
{\tilde\sigma}_{(q{\bar q})p}({l}_{\perp}^{2}) &=&
\frac{{\sigma}_{(q{\bar q})p}^{(\infty)}}{\pi}{R}_{0}^{2}
{e}^{-{l}_{\perp}^{2}R_0^2} \quad  \Rightarrow \quad 
{\sigma}_{(q{\bar q})p}({r}_{\perp}^{2})={\sigma}_{(q{\bar q})p}^{(\infty)}
\left(1-{e}^{-\frac{{r}_{\perp}^{2}}{4{R}_{0}^{2}}}\right) \ .
\label{gauss}
\end{eqnarray}
For simplicity of notation, in (\ref{delta}) and (\ref{gauss})
 the $W^2$--dependence
of ${\tilde\sigma}_{(q{\bar q})p}(l^2_{\perp})$,
${\sigma}_{(q{\bar q})p}(r^2_{\perp})$ and
${\sigma}_{(q{\bar q})p}^{(\infty)}$ was dropped.
From the subsequent examination of the transverse and the longitudinal cross
 section in (\ref{ip4}), not unexpectedly, one finds that (\ref{delta}) and
 (\ref{gauss}) lead to approximately the same results, provided one identifies
the parameters $\Lambda$ and $R_0$ via $\Lambda = 1/R_0$, where $R_0$ is 
of the order of the proton radius, $R_0 \approx 1$ fm $ \approx 0.2 \  
{\rm GeV}^{-1}$. 
We note that a Gaussian ansatz was employed in a recent analysis
\cite{Wuesthof} of the experimental data. A different, polynomial 
representation for the $r_{\perp}^2$--dependence of the
 dipole cross section is used in Ref.~\cite{Shaw}. 
\newpage
\begin{figure}[htb]
\setlength{\unitlength}{1.cm}
\begin{center}
\epsfig{file=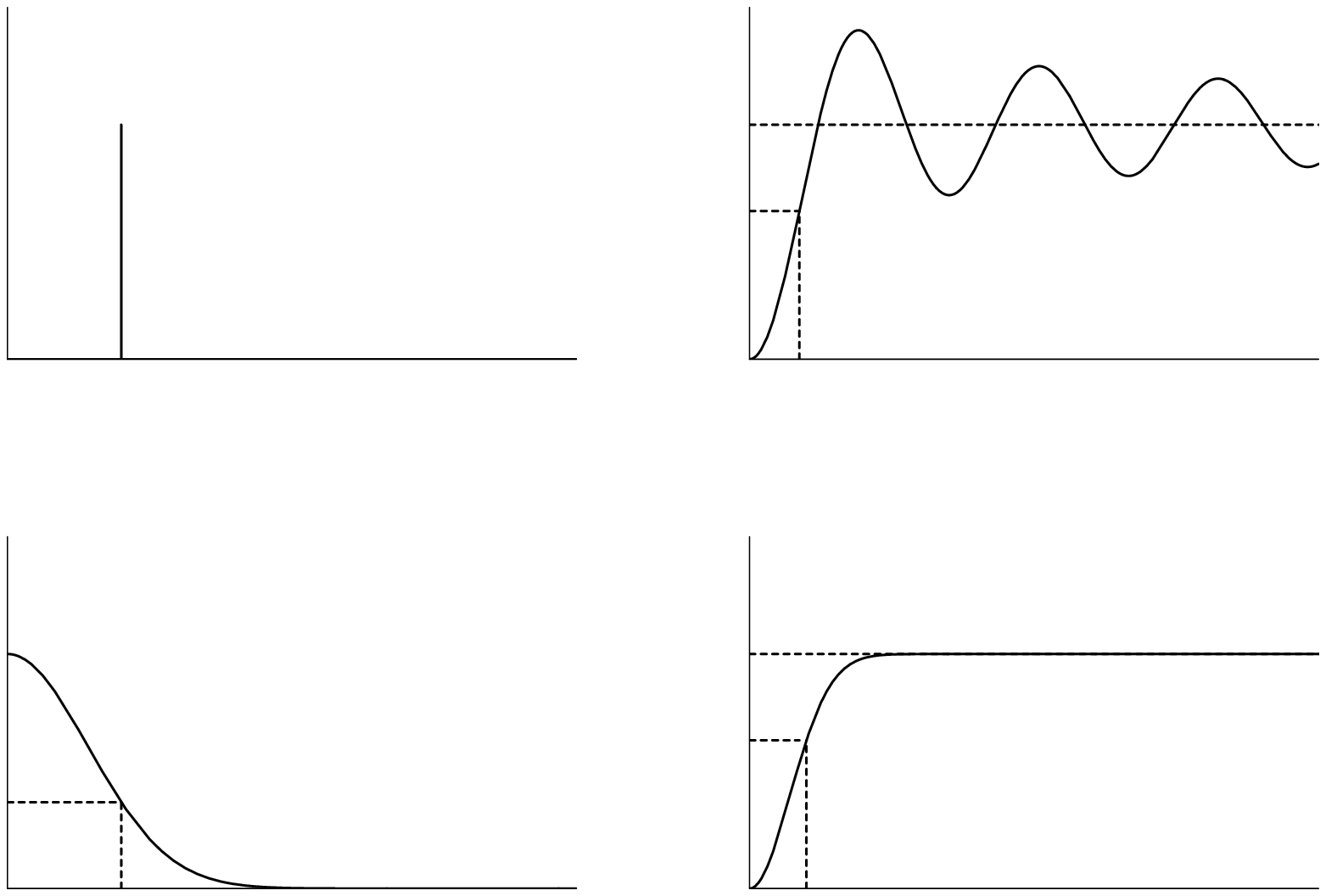, height=16cm, width=16cm}
\put(-1,1.2){\makebox(1,1)[c]{${r}_{\perp}$}}
\put(-1,9.2){\makebox(1,1)[c]{${r}_{\perp}$}}
\put(-9.5,1.2){\makebox(1,1)[c]{${l}_{\perp}$}}
\put(-9.5,9.2){\makebox(1,1)[c]{${l}_{\perp}$}}
\put(-15.7,7.2){\makebox(1,1)[c]{${{\tilde \sigma}_{(q{\bar q})p}({l}_{\perp}^2,W^2)}$}}
\put(-7.2,7.2){\makebox(1,1)[c]{${{\sigma}_{(q{\bar q})p}({r}_{\perp}^2,W^2)}$}}
\put(-7.2,15.2){\makebox(1,1)[c]{${{\sigma}_{(q{\bar q})p}({r}_{\perp}^2,W^2)}$}}
\put(-15.7,15.2){\makebox(1,1)[c]{${{\tilde \sigma}_{(q{\bar q})p}({l}_{\perp}^2,W^2)}$}}
{\footnotesize
\put(-16.4,2.9){\makebox(1,1)[c]{$\frac{{\tilde \sigma}(0)}{e}$}}
\put(-8,5.2){\makebox(1,1)[c]{$\sigma_{(q{\bar q})p}^{(\infty)}$}}
\put(-8,13.1){\makebox(1,1)[c]{$\sigma_{(q{\bar q})p}^{(\infty)}$}}
\put(-14.75,9.2){\makebox(1,1)[c]{$l_{\perp}\!=\!\Lambda$}}
\put(-14.55,1.2){\makebox(1,1)[c]{$l_{\perp}\!=\!1/R_0$}}
\put(-6.55,1.2){\makebox(1,1)[c]{$r_{\perp}\!=\!2 \cdot R_0$}}
\put(-6.4,9.2){\makebox(1,1)[c]{$r_{\perp}\!=\!1.75/\Lambda$}}
\put(-8.4,3.8){\makebox(1,1)[c]{$0.63 \cdot {\sigma}_{(q{\bar q})p}^{(\infty)}$}}
\put(-8.4,11.8){\makebox(1,1)[c]{$0.63\cdot{\sigma}_{(q{\bar q})p}^{(\infty)}$}}}
\put(-8.7,8.3){\makebox(1,1)[c]{${\small a)}$}}
\put(-8.7,0.5){\makebox(1,1)[c]{${\small b)}$}}
\end{center}
\caption{\small The transverse-position-space dipole cross section
  ${\sigma}_{(q{\bar q})p}({r}^2_{\perp},W^2)$ and its Fourier transform
${\tilde \sigma}_{(q{\bar q})p}({l}^2_{\perp},W^2)$ for two simple choices in
transverse momentum space,  
a) for a ${\delta}$--function and b) for a Gaussian.}
\label{Fig4}
\end{figure}
\clearpage

\section{Evaluation of ${\sigma}_{{\gamma}^{\ast}_{T, L}p}(W^2, Q^2)$, the
  explicit connection with off-diagonal generalised vector dominance.}
\label{section4}

The dependence of the ${\gamma}^{\ast}\!\to\!q{\bar q}$ 
transition amplitudes (\ref{j2}) and (\ref{j3}) 
on the propagator of the $q{\bar q}$ system 
of mass ${M}_{q{\bar q}}$ suggests a change of the integration
variables in ${\sigma}_{{\gamma}_{T,L}^{\ast}p}$ in the expression (\ref{sigtot1}). The angular
integration over the direction of the transverse momentum of the incoming 
quark, ${\vec k}_{\perp}$, yields a factor $2 \pi$, and we end up with
\begin{eqnarray}
\int_0^1\!dz &&\int_0^{\infty}\!d^2{\vec l}_{\perp}\int_{k_{\perp
    0}}^{\infty}\!d^2{\vec k}_{\perp}\ldots 
\nonumber\\
&&
={\pi}\int_0^1\!dz z(1\!-\!z)\!\int_{0}^{\infty}\!d{l}^2_{\perp}
\int_{M_0^2(z)}^{\infty}\!dM^2
\int_{(M-l_{\perp}^{\prime}(z))^2}^{(M+l_{\perp}^{\prime}(z))^2}\! 
dM^{\prime 2} w(M^2, M^{\prime 2}, {l}_{\perp}^{\prime 2}(z))\ldots \ ,
\label{trans}
\end{eqnarray}
where 
\begin{equation}
 M_0^2(z)= \frac{k^2_{\perp 0}}{z(1-z)} \ .
\label{M0z}
\end{equation}
In (\ref{trans}), we omitted the subscripts
$q{\bar q}$ at the squared masses $M^2$ (\ref{M}) and $M^{\prime 2}$ (\ref{Mp}).
The weight function $w(M^2, M^{\prime 2}, {l}_{\perp}^{2}, z)$ 
appearing in (\ref{trans}) is given by
\begin{equation}
w(M^2, M^{\prime 2}, {l}_{\perp}^{\prime 2}(z))= 
\frac{1}{2 M {M}^{\prime} \sqrt{1-\cos^2\phi}} \ .
\label{weightf}
\end{equation}
The angle between ${\vec k}_{\perp}$ and 
${\vec k}_{\perp}\!+\!{\vec l}_{\perp}$ has been denoted by $\phi$
(cf.~Fig.~5) and $\cos^2 \phi$, as a function of $M^2$, $M^{\prime 2}$, 
${l}_{\perp}^{2}$ and $z$, is constrained by
\begin{equation}
\cos^2 \phi \equiv \frac{1}{4 M^2 {M}^{\prime 2}} 
\left ({M}^{2}\!+\!M^{\prime 2}\!-\!\frac{{l}_{\perp}^2}{z (1\!-\!z)}\right )^2 
\leq 1 \ .
\label{cosq}
\end{equation}
\begin{figure}[htb]
\setlength{\unitlength}{1.cm}
\begin{center}
\epsfig{file=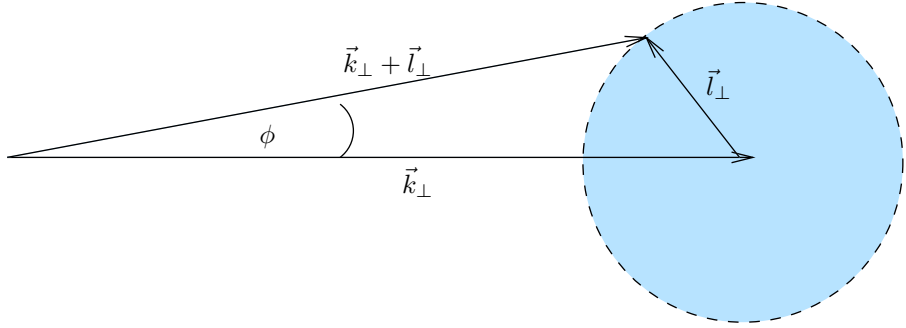, width=12cm}
\put(-3,2.7){\makebox(1,1)[c]{\footnotesize ${\vec l}_{\perp}$}}
\put(-7,1.4){\makebox(1,1)[c]{\footnotesize ${\vec k}_{\perp}$}}
\put(-7.4,3){\makebox(1,1)[c]{\footnotesize ${\vec k}_{\perp}+{\vec l}_{\perp}$}}
\put(-9,2){\makebox(1,1)[c]{\footnotesize $\phi$}}
\end{center}
\caption{The pictorial expose of the quantities ${\vec k}_{\perp}$, ${\vec
    l}_{\perp}$, ${\vec k}_{\perp}+{\vec l}_{\perp}$, and the angle $\phi$.}
\label{Fig5}
\end{figure}
This constraint implies bounds on the integration interval for the integration
over $d {M}^{\prime 2}$. As indicated in (\ref{trans}), the bounds are given by
${(M\!\pm\!{l}^{\prime}_{\perp}(z))}^2$, where
\begin{equation}
{l}^{\prime 2}_{\perp} \equiv {l}^{\prime 2}_{\perp}(z)=\frac{{l}^2_{\perp}}{z (1\!-\!z)} \ .
\label{ldef}
\end{equation}
For later use we note
\begin{equation}
\int_{ (M-{l}_{\perp}^{\prime}(z))^2 }^{ (M+{l}_{\perp}^{\prime}(z))^2 }
d M^{\prime 2} w(M^2, M^{\prime 2}, {l}_{\perp}^{\prime 2}(z)) = \pi \ ,
\label{intm}
\end{equation}
as well as 
\begin{equation}
\frac{\int_{ (M-{l}_{\perp}^{\prime}(z))^2 }^{ (M+{l}_{\perp}^{\prime}(z))^2 }
d M^{\prime 2} w(M^{2}, M^{\prime 2}, {l}_{\perp}^{\prime 2}(z)) M^{\prime 2}}
{\int_{ (M-{l}_{\perp}^{\prime}(z))^2 }^{ (M+{l}_{\perp}^{\prime}(z))^2 }
d M^{\prime 2} w(M^2, M^{\prime 2}, {l}_{\perp}^{\prime 2}(z))} = 
(M^2 + {l}_{\perp}^{\prime 2}(z)) \ . 
\label{intm1}
\end{equation}
In order to clarify the physical meaning of the right-hand side of (\ref{intm1}), we
observe that (\ref{sigtot1}), upon removal of the $Q^2$--dependent propagator terms,
becomes proportional to the purely hadronic cross section 
${\tilde \sigma}_{(q{\bar q})p}$. Consequently, the quantity 
in (\ref{intm1}) is the mean  mass produced in the 
$(q{\bar q})p$ forward-scattering reaction at fixed values of $M$, $z$ and 
$l_{\perp}$.

In passing from the integration variables in (\ref{sigtot1}) to the new ones
introduced in (\ref{trans}), the integration limits on the integration over 
$dM^{\prime 2}$ and $dM^{2}$ have to be carefully looked at.

In the integration over $dM^{\prime 2}$, we first consider the first 
term in the curly brackets of (\ref{sigtot1}), i.e.,
the term diagonal in the mass $M\!\equiv\!M_{q{\bar q}}$ of the
$q{\bar q}$ pair. In this term, the integration over
$dM^{\prime 2}$ [cf.~(\ref{Mp})] corresponds to 
an integration over all directions of ${\vec l}_{\perp}$ in 
Fig.~\ref{Fig5}, i.e. to an integration over the range of
$\phi$ allowed by (\ref{cosq}) at fixed values of $M^2$, $l_{\perp}^2$ and
$z$. The corresponding integration limits are as indicated in (\ref{trans}).

Concerning the second term, i.e. the off-diagonal one in the curly brackets of 
(\ref{sigtot1}), we note that in this term integration over $dM^{\prime 2}$
describes integration over all final-state 
$q{\bar q}$ masses (\ref{Mp}) in the Compton
forward amplitude. For the off-diagonal term, consequently, the lower limit of
integration in (\ref{trans}), namely $(M\!-\!l^{\prime}_{\perp}(z))^2$, must 
in addition be restricted to values above 
$M_0^2(z)$ from (\ref{M0z}), as indicated in ({\ref{sigtot1}) already. 
After all, the photon couplings to the
initial and final $q{\bar q}$ state in the Compton forward amplitude must be 
identical. 

We turn to the integration over $dM^2$ in (\ref{trans}). 
In the diagonal term in (\ref{sigtot1}), it describes integration over the ingoing
and the outgoing mass, while in the off-diagonal term it describes integration
over the mass of the incoming  $q{\bar q}$ pair only. 
The necessity of a lower cutoff 
$M_0^2$ in the integration over $dM^2$ stems from the empirical fact that 
$e^+e^- \to {\gamma}^{\ast} \to q{\bar q} \to {\rm hadrons}$ becomes appreciable only 
when the centre-of-mass energy $\sqrt{s(e^+ e^-)}$ is above a
lower threshold that depends on the flavour of the quark $q$. This  
suggests a limit of  $M_0 \leq M_{\rho}$, for $u$ and $d$ quarks, while $M_0
\leq M_{J/{\Psi}}$ for the $c$ quarks, etc. Actually, in (\ref{trans}), we have
indicated a $z$--dependent lower cutoff of
$M_0^2(z)\!=\!k_{\perp 0}^2/(z(1\!-\!z))$.
It originates from the lower bound on the transverse momentum of the quark 
$q$ in the incoming $q{\bar q}$ state as introduced in (\ref{fj1}). While 
this $z$--dependent bound on squared masses $M^2_{q{\bar q}}$
is thus suggested by confinement, it should be kept
in mind that the description of the coupling of the photon to the low--lying
resonances in terms of a simple ${\gamma}^{\ast} \to q{\bar q}$ transition amplitude
is an effective one \cite{SakuraiSchildknecht} in the sense of 
averaging the total cross section over the contributing resonances,
e.g., the $\rho^0$, $\rho^{\prime 0}$, etc\footnote{A two-component ansatz
  (low-mass vector mesons plus high-mass $q{\bar q}$ jets) is frequently 
employed \cite{Badelev}. We believe that an effective single-component picture
\cite{SakuraiSchildknecht} will be sufficiently accurate.}. The $z$ dependence of the lower
bound, $M_0^2(z)$, is accordingly to be looked at with some reservation. 
We will comment on the effect of the $z$ dependence of $M_0^2$ in the
numerical analysis of Section \ref{section5}.

Returning to (\ref{sigtot1}), 
inserting the expressions (\ref{j2}) and (\ref{j3})
for the ${\gamma}^{\ast} \to q{\bar q}$ transitions,
and introducing the integration variables $M^2$ and $M^{\prime 2}$ 
according to  (\ref{trans}),
we get for the transverse photoabsorption cross section 
\begin{eqnarray}
&&{\sigma}_{ {\gamma}^{\ast}_T p }(W^2, Q^2) 
 = \frac{\alpha}{2 {\pi}} \left( \frac{e_q}{e_0} \right)^2  
\int_{0}^{\infty} d {l}^2_{\perp} 
{\tilde \sigma}_{(q{\bar q})p} \left({l}_{\perp}^{2},W^2 \right)
\int_0^1 d z (1\!-\!2z(1\!-\!z))
\times
\nonumber\\
&&
{\Bigg \lbrace}
\int_{M_0^2(z)}^{\infty} d M^2 
\int_{(M-{l}_{\perp}^{\prime}(z))^2}^{(M+{l}_{\perp}^{\prime}(z))^2} d M^{\prime 2} 
w(M^{2}, M^{\prime 2}, {l}_{\perp}^{\prime 2}(z))
\left \lbrack \frac{M^2}{(Q^2+M^2)^2}
-\frac{M^{\prime 2}+M^2-{l}_{\perp}^{\prime 2}(z)}{2 (Q^2+M^2)(Q^2+{M}^{\prime
 2})}\right \rbrack 
\nonumber\\
&&
+ \int_{M_0^2(z)}^{\infty}\!\!dM^2 
\Theta(M_0^2(z)\!-\!(M\!-\!{l}_{\perp}^{\prime}(z))^2) \times
\nonumber\\
&&
\int_{(M\!-\!{l}_{\perp}^{\prime}(z))^2}^{M_0^2(z)}\!dM^{\prime 2}w(M^{2},
M^{\prime 2}, {l}_{\perp}^{\prime 2}(z))
\frac{M^{\prime 2}\!+\!M^2\!-\!{l}_{\perp}^{\prime 2}(z)}
{2 (Q^2\!+\!M^2)(Q^2\!+\!{M}^{\prime 2})}
{\Bigg \rbrace} \ , \nonumber\\
\label{sigTA}
\end{eqnarray}
and for the longitudinal one 
\begin{eqnarray}
&&{\sigma}_{ {\gamma}^{\ast}_L p }(W^2, Q^2) 
 = \frac{2 \alpha}{{\pi}} 
\left( \frac{e_q}{e_0} \right)^2 Q^2
\int_{0}^{\infty} d {l}^2_{\perp} 
{\tilde \sigma}_{(q{\bar q})p} \left({l}_{\perp}^{2},W^2 \right)
\int_0^1 d z z (1\!-\!z)
\times 
\nonumber\\
&&
{\Bigg \lbrace}
\int_{M_0^2(z)}^{\infty} d M^2
\int_{(M-{l}_{\perp}^{\prime}(z))^2}^{(M+{l}_{\perp}^{\prime}(z))^2} d
M^{\prime 2} w(M^{2}, M^{\prime 2}, {l}_{\perp}^{\prime 2}(z)){\Bigg \lbrack}
\frac{1}{(Q^2+M^2)^2} -
\frac{1}{(Q^2+M^2)(Q^2+{M}^{\prime 2})}{\Bigg \rbrack}
\nonumber\\
&&
+ \int_{M_0^2(z)}^{\infty}\!\!dM^2 
\Theta(M_0^2(z)\!-\!(M\!-\!{l}_{\perp}^{\prime}(z))^2) \times
\nonumber\\
&&
\int_{(M\!-\!{l}_{\perp}^{\prime}(z))^2}^{M_0^2(z)}\!dM^{\prime 2} 
w(M^{2},M^{\prime 2}, {l}_{\perp}^{\prime 2}(z)) 
\frac{1}{(Q^2\!+\!M^2)(Q^2\!+\!{M}^{\prime 2})}
{\Bigg \rbrace} \ ,
\label{sigLA}
\end{eqnarray} 
where $\Theta(x)$ is the step function [$\Theta(x)\!=\!1$ for $x\!>\!0$,
$\Theta(x)\!=\!0$ otherwise].
In (\ref{sigTA}) and (\ref{sigLA}), the $\Theta$--function term becomes
unequal zero as soon as 
$(M\!-\!l_{\perp}^{\prime}(z))^2$ drops below $M_0^2(z)$, thus
removing the above-mentioned forbidden region 
in the integration over $dM^{\prime 2}$ in the (main) off-diagonal term.
We note that the ``low-mass term'' containing the $\Theta$--function is 
suppressed relative to the main off-diagonal term, as the intervals of the
integration over $dM^2$ and $dM^{\prime 2}$
are very much restricted, 
\begin{eqnarray}
\lefteqn{
\int_{M_0^2(z)}^{\infty}\!dM^2 
\Theta \left( M_0^2(z)\!-\!(M\!-\!{l}_{\perp}^{\prime}(z))^2 \right)
\int_{(M\!-\!{l}_{\perp}^{\prime}(z))^2}^{M_0^2(z)}
d M^{\prime 2} \ldots }
\nonumber\\
&& 
= \int_{M_0^2(z)}^{(M_0(z)\!+\!{l}_{\perp}^{\prime}(z))^2}\!dM^2\!
\int_{(M\!-\!{l}_{\perp}^{\prime}(z))^2}^{M_0^2(z)}\!dM^{\prime 2} \dots \ .
\label{bounds}
\end{eqnarray}
Actually, it will turn out that the main term in the transverse cross section
will asymptotically behave like $(\ln Q^2)/Q^2$, thus suppressing the 
${\Theta}$--function term that behaves as $1/Q^2$. In the longitudinal cross
section the suppression is less pronounced, as both the main term and the 
${\Theta}$--function term behave as $1/Q^2$ for asymptotic $Q^2$. The 
subsequent analysis of this Section will be simplified by ignoring the 
${\Theta}$--function terms in (\ref{sigTA}) and (\ref{sigLA}). We will come 
back to them, when turning to the numerical results in Section \ref{section5}.

In order to explicitly obtain the $Q^2$ dependence of
${\sigma}_{{\gamma}^{\ast}_Tp}$ and 
${\sigma}_{{\gamma}^{\ast}_Lp}$ contained in (\ref{sigTA})
and (\ref{sigLA}), we proceed in several steps. In a first step, we will show
that the transverse cross section (\ref{sigTA}) may be evaluated analytically in the limit of
$Q^2 \to \infty$ for the simple case of the $\delta$--function ansatz (\ref{delta}) for
${\tilde \sigma}_{(q{\bar q})p}(l^2_{\perp}, W^2)$. In the second step, we
introduce mean values for the configuration variable, $z$, and for $M^{\prime 2}$ at
fixed $M^2$ and $l^2_{\perp}=\Lambda^2$, 
and apply the mean-value theorem to the integration over $z$ and $M^{\prime 2}$ 
in the transverse cross
section (\ref{sigTA}) for arbitrary values of  $Q^2$. A similar
procedure will be carried out for the longitudinal cross section.
After these steps, the connection with
the original formulation of off-diagonal GVD \cite{FRS} will become
explicit. 
The determination of the numerical values of the mean configuration variables,
${\bar z}_{T,L}$, and of the parameters $\delta_{T,L}$ characterising the mean 
mass, $M^{\prime}$, will be shifted to
Section \ref{section5}.

\subsection{The transverse cross section, ${\sigma}_{{\gamma}^{\ast}_T p}$.}

As noted, we ignore the ${\Theta}$--function term in (\ref{sigTA}),
insert the $\delta$--function ansatz (\ref{delta}) for ${\tilde
  \sigma}_{q{\bar q}p}$, and carry out the 
integrations over $dl_{\perp}^2$ and $dM^{\prime 2}$, to obtain
\begin{eqnarray}
{\sigma}_{ {\gamma}^{\ast}_T p }&(&W^2, Q^2) 
 = \frac{\alpha}{2 {\pi}} \left( \frac{e_q}{e_0} \right)^2 
{\sigma}^{(\infty)}_{(q{\bar q})p}
\int_0^1 d z (1\!-\!2z(1\!-\!z))
\int_{M_0^2(z)}^{\infty} d M^2 
\times
\nonumber\\
&&
\left \lbrack \frac{M^2}{(Q^2+M^2)^2}
-\frac{1}{(Q^2+M^2)} \frac{1}{2} \left (1+\frac{M^2-Q^2-{\Lambda}^{\prime 2}}
{\sqrt{(Q^2+M^2+{\Lambda}^{\prime 2})^2-4 {\Lambda}^{\prime 2} M^2}}\right
 ) \right \rbrack\ ,
\label{sigTA5}
\end{eqnarray}
where ${\Lambda}^{\prime 2}\!\equiv\!{\Lambda}^2/(z(1\!-\!z))$.
Replacing $z$ by the variable $u \equiv z(1-z)$ yields
\begin{eqnarray}
{\sigma}_{ {\gamma}^{\ast}_T p }&(&W^2, Q^2) 
 = \frac{{\alpha}}{\pi}\left( \frac{e_q}{e_0} \right)^2 
{\sigma}^{(\infty)}_{(q{\bar q})p}
\int_0^{1/4} d u \frac{1-2u}{\sqrt{1-4u}}
\int_{M_0^2(u)}^{\infty} d M^2 
\times
\nonumber\\
&&
\left \lbrack \frac{M^2}{(Q^2+M^2)^2}
-\frac{1}{(Q^2+M^2)} \frac{1}{2} \left (1+\frac{u(M^2-Q^2)-{\Lambda}^{2}}{
  \sqrt{(u(Q^2+M^2)+{\Lambda}^{2})^2-4 u {\Lambda}^{2} M^2}}\right
 )\right \rbrack \ .
\label{sigTA6}
\end{eqnarray}
Expansion of the off-diagonal term in a power series for large $Q^2$ gives
\begin{eqnarray}
&&{\sigma}_{ {\gamma}^{\ast}_T p }(W^2, Q^2) 
 = \frac{{\alpha}}{\pi}
\left( \frac{e_q}{e_0} \right)^2 
{\sigma}^{(\infty)}_{(q{\bar q})p}
\int_0^{1/4} d u \frac{1-2u}{\sqrt{1-4u}}
\int_{M_0^2(u)}^{\infty} d M^2 
\frac{ \Lambda^2 }{(Q^2 + M^2) (u (Q^2\!+\!M^2) + \Lambda^2)} \times
\nonumber\\
&&
\left[ 
\frac{M^2}{(Q^2+M^2)} + \frac{u M^2}{ (u (Q^2\!+\!M^2) + \Lambda^2) }
- \frac{2 u^2 M^4}{ (u (Q^2\!+\!M^2) + \Lambda^2)^2 } + \cdots 
\right] \ .
\label{sigTA6a}
\end{eqnarray}
The integration over $dM^2$ and $du$ in (\ref{sigTA6a}) may now be carried out 
by expanding $1/\sqrt{1\!-\!4u}$ in powers of $u$
and integrating term by term. 
It turns out that the replacement of $1/\sqrt{1\!-\!4u}$
by $1\!+\!2u$ is sufficient to yield the leading term in 
the large--$Q^2$ limit of $Q^2\!\gg\!{\Lambda}^2$,
\begin{equation}
{\sigma}_{ {\gamma}^{\ast}_T p }(W^2, Q^2) 
 = \frac{{\alpha}}{3 \pi}\left( \frac{e_q}{e_0} \right)^2
{\sigma}^{(\infty)}_{(q{\bar q})p} \left \lbrack 
\frac{{\Lambda}^2}{Q^2} \ln \left (\frac{Q^2}{{\Lambda}^2} \right )+ 
c \frac{\Lambda^2}{Q^2} + {\cal O} \left ( \frac{\ln Q^2}{Q^4} \right ) 
\right \rbrack  \ .
\label{sigTqaq}
\end{equation}
We note that the leading term in (\ref{sigTqaq}) is independent of the 
threshold mass parameterised by $k_{\perp 0}^2$. Moreover, (\ref{sigTqaq})
shows a logarithmic violation of scaling of the transverse part of the 
structure function $F_2\!\sim\!Q^2 {\sigma}_{{\gamma}^{\ast}p}$. The constant 
$c$ in (\ref{sigTqaq}) is a complicated function of $\Lambda^2/k^2_{\perp
  0}$. For $\Lambda^2/k^2_{\perp 0}\!\sim\!1$ its value is $c\!\sim\!1$. We
remark at this point that the $\Theta$--function term ignored so far, according
to the numerical analysis of Section \ref{section5}, will modify the numerical
values of $c$ only, while leaving the leading term in (\ref{sigTqaq}) unchanged.

We turn to the second step in the evaluation of the transverse cross section.
The use of the mean-value theorem removes the integral over $dz$ in
(\ref{sigTA}), $z$ being replaced by its mean value, ${\bar z}_T$, and,
accordingly, $l_{\perp}^{\prime 2}(z)$ by 
\begin{equation}
{\bar l}^{\prime 2}_{\perp} \equiv l^{\prime 2}_{\perp}
({\bar z}_T)=\frac{l_{\perp}^2}{{\bar z}_T(1-{\bar z}_T)} \ .
\label{lmitt}
\end{equation}
With respect to the integration over $dM^{\prime 2}$, we note that in the 
large--$Q^2$ limit the $M^{\prime 2}$--dependent propagator 
part in (\ref{sigTA}), is given by
\begin{equation}
\frac{M^{\prime 2}}{(Q^2+M^{\prime 2})} = \frac{M^{\prime 2}}{Q^2}-
\frac{M^{\prime 4}}{Q^4}+{\cal O} \left (\frac{M^{\prime 6}}{Q^6} \right ) \ .
\label{aprox}
\end{equation}
As far as the first term on the right-hand side of (\ref{aprox}) is concerned,
integration over $dM^{\prime 2}$ in the off-diagonal term in (\ref{sigTA}), 
according to (\ref{intm1}), corresponds to replacing $M^{\prime 2}$ 
by $\!M^2\!+\!{\bar l}_{\perp}^{\prime 2}$ 
when applying the mean-value theorem. As (\ref{sigTA}) contains the full
left-hand side of (\ref{aprox}), the mean value of $M^{\prime 2}$ will deviate
from $\!M^2\!+\!{\bar l}_{\perp}^{\prime 2}$, in particular for small values of
$Q^2$. Accordingly, we introduce the parameter $\delta_T$ to express the mean
value, ${\overline M^{\prime 2}}$, of $M^{\prime 2}$ in terms of $M^2$ and
${\bar l}_{\perp}^{\prime 2}$ by 
\begin{equation}
{\overline M^{\prime 2}}= M^2+
\frac{{\bar l}_{\perp}^{\prime 2}}{(1\!+\!2 {\delta}_T)} \ .
\label{Mmitt1}
\end{equation}
After these steps, we have
\begin{eqnarray}
{\sigma}_{ {\gamma}^{\ast}_T p }(W^2, Q^2; {\bar z}_T, \delta_T) 
& = & \frac{\alpha}{2} \left( \frac{e_q}{e_0}
\right)^2 (1-2{\bar z}_T(1\!-\!{\bar
  z}_T)) \int_0^{\infty}dl_{\perp}^2 {\tilde \sigma}_{(q{\bar q})p}(l^2_{\perp},W^2)
\int_{M_0^2({\bar z})}^{\infty} d M^2 \times 
\nonumber\\
&&
\left \lbrack \frac{M^2}{(Q^2+M^2)^2} 
- \frac{ M^2 - {\bar l}_{\perp}^{\prime 2} \delta_T / (1\!+\!2\delta_T) }
{(Q^2+M^2) (Q^2+M^2+{\bar l}_{\perp}^{\prime 2} / (1\!+\!2\delta_T) )}
\right \rbrack \ .
\label{sigTA11}
\end{eqnarray}
Upon inserting the $\delta$--function ansatz (\ref{delta})
for ${\tilde \sigma}_{(q{\bar q})p}$, the cross section,\footnote{
In analogy with (\ref{lmitt}), we denote here
${\bar \Lambda}^{\prime 2} \equiv \Lambda^2/({\bar z}_T(1\!-\!{\bar z}_T))$.} 
\begin{eqnarray}
{\sigma}_{ {\gamma}^{\ast}_T p }(W^2, Q^2; {\bar z}_T, \delta_T) 
& = & \frac{\alpha}{2 \pi} \left( \frac{e_q}{e_0}
\right)^2 {\sigma}^{(\infty)}_{(q{\bar q})p}(1-2{\bar z}_T(1\!-\!{\bar
  z}_T)) 
\int_{M_0^2({\bar z})}^{\infty} d M^2 \times 
\nonumber\\
&&
\left \lbrack \frac{M^2}{(Q^2+M^2)^2} - 
\frac{M^2 - {\bar \Lambda}^{\prime 2} \delta_T / (1\!+\!2\delta_T) }
{(Q^2+M^2) (Q^2+M^2 + {\bar \Lambda}^{\prime 2} / (1\!+\!2\delta_T) ) }
\right \rbrack \ ,
\label{sigTA1}
\end{eqnarray}
explicitly coincides with the continuum version\footnote{Indeed, the expression
(4) of Ref.\cite{FRS} upon substitution of (6) of Ref.\cite{FRS} agrees with
(\ref{sigTA1}) upon identification of $\lambda m_0^2$ with $\lambda m_0^2
\equiv {\bar \Lambda}^{\prime 2}/(1+2\delta_T)$.} of off-diagonal GVD
\cite{FRS}. The original ansatz of off-diagonal GVD has thus been recovered
from the QCD-motivated ansatz
(\ref{sigtot1}) by
introducing mean values for the configuration variable, ${\bar z}_T$, and 
for the outgoing mass $M^{\prime 2}$ via $\delta_T$.

Carrying out the remaining integration over $dM^2$ in (\ref{sigTA1}), we have
\begin{eqnarray}
{\sigma}_{ {\gamma}^{\ast}_T p }&(&W^2, Q^2; {\bar z}_T, \delta_T) 
= \frac{\alpha}{2 \pi} \left( \frac{e_q}{e_0}
\right)^2 {\sigma}^{(\infty)}_{(q{\bar q})p}(1-2{\bar z}_T(1\!-\!{\bar z}_T))
\times
\nonumber\\
&&
{\Bigg \lbrack} 
\left (
(1\!+\!2 \delta_T) \frac{Q^2}{{\bar \Lambda}^{\prime 2}}+(1\!+\!\delta_T)\right)
\ln \left (1+\frac{{\bar \Lambda}^{\prime 2}}{(1\!+\!2 \delta_T)
(Q^2+M_0^2({\bar z}_T))}\right )
- \frac{Q^2}{ (Q^2+M_0^2({\bar z}_T)) } 
{\Bigg \rbrack} \ .
\label{sigTA2}
\end{eqnarray} 
The numerical results for the mean values of 
${\bar z}_T$ and of $\delta_T$
may be determined by comparing with a numerical evaluation of
(\ref{sigTA}). It is suggestive, to determine ${\bar z}_T$ and $\delta_T$ 
at the fixed value of $Q^2=0$ by adjusting the photoproduction limit of (\ref{sigTA2}),  
\begin{equation}
{\sigma}_{ {\gamma}p }(W^2; {\bar z}_T, \delta_T) 
= \frac{\alpha}{2 \pi} \left( \frac{e_q}{e_0}
\right)^2 {\sigma}^{(\infty)}_{(q{\bar q})p}(1-2{\bar z}_T(1\!-\!{\bar z}_T)) 
(1\!+\!\delta_T) \ln 
\left (1+\frac{{\Lambda}^{2}}
{(1\!+\!2 \delta_T) k_{\perp 0}^2}\right ) \ , 
\label{sigTA4}
\end{equation}
and the derivative of ${\sigma}_{ {\gamma}^{\ast}_Tp }$  
with respect to $Q^2$ at $Q^2=0$ to the corresponding numerical results from
(\ref{sigTA}). Details will be presented  in Section \ref{section5}. We only
note the results of 
\begin{equation}
\kappa_{T}(0) \equiv {\bar z}_T(1-{\bar z}_T)=0.1455, \quad
\delta_T=0.5224
\label{kappa0}
\end{equation}
for the choice of $\Lambda^2/k_{\perp 0}^2=1$ that will be
adopted as a preferred one. In (\ref{kappa0}), the notation $\kappa_T(0)$ is 
introduced to indicate that $\kappa_T$ is determined at $Q^2=0$.

Taking the large--$Q^2$ limit of (\ref{sigTA2}), we find 
\begin{equation}
{\sigma}_{ {\gamma}^{\ast}_T p }(W^2, Q^2 \to \infty; {\bar z}_T) 
= \frac{\alpha}{3 \pi} \left( \frac{e_q}{e_0} \right)^2 
{\sigma}^{(\infty)}_{(q{\bar q})p} \frac{3}{4}
\frac{(1-2{\bar z}_T(1\!-\!{\bar z}_T))}{ {\bar z}_T(1-{\bar z}_T)}
\left \lbrack \frac{{\Lambda}^{2}}{Q^2} + {\cal O}\left
 (\frac{1}{Q^4}\right )
\right \rbrack \ .
\label{sigTA3}
\end{equation}
The dependence on $\delta_T$ in (\ref{sigTA2}) has dropped out for $Q^2 \to
\infty$. This is as expected, when taking into account (\ref{aprox}) and (\ref{intm1}).
A comparison of (\ref{sigTA3}) with the exact large--$Q^2$ limit in 
(\ref{sigTqaq}) reveals that the application of the mean-value theorem
suppresses the $\ln Q^2$ factor of the transverse cross section that is present
according to (\ref{sigTqaq}), whereas the $1/Q^2$ behaviour
relevant for scaling of the structure function $F_2$ remains. The loss of 
the $\ln Q^2$ factor may uniquely be traced back to the introduction of 
${\bar z}_T$; in fact, introducing ${\bar z}_T$ in (\ref{sigTA}), but carrying
out the integration over $dM^{\prime 2}$ analytically, as in (\ref{sigTA5}), 
the $\ln Q^2$ term is lost as well. This suggests that the 
appearance of the configuration variable $z$ in the integrand of
(\ref{sigTA}) is irrelevant for the  $1/Q^2$ (scaling) behaviour. It is
responsible, however, for the logarithmic violation of scaling. 
Effectively, ${\bar z}_T$, the mean value of $z$ that determines the cross section, changes
with increasing $Q^2$, thus leading to the additional $\ln Q^2$ dependence in
(\ref{sigTqaq}).
This will be shown explicitly by introducing a $Q^2$ dependence for 
${\bar z}_T(1-{\bar z}_T)$ in (\ref{sigTA2}) that reproduces 
${\sigma}_{ {\gamma}^{\ast}_T p }$ from (\ref{sigTA}) with its correct
asymptotic behaviour (\ref{sigTqaq}).

We proceed in two steps. In a first step, we note that the ratio 
\begin{equation}
r_{T}(Q^2) \equiv
\frac{{\sigma}_{{\gamma}_{T}^{\ast}p(W^2, Q^2)}}
{{\sigma}_{{\gamma}_{T}^{\ast}p(W^2, Q^2; {\bar z}_T, \delta_T)}} \ ,
\label{gerh}
\end{equation}
as a consequence of the above-mentioned determination of ${\bar z}_T$ and
$\delta_T$ at $Q^2=0$, fulfills
\begin{equation}
r_T(Q^2=0)=1 \ .
\label{rQ0}
\end{equation}
For $Q^2 \to \infty$, according to (\ref{sigTqaq}) and (\ref{sigTA3}), on the
other hand, we have
\begin{equation}
r_T(Q^2 \to \infty)=\frac{4\kappa_T(0)}{3(1-2\kappa_T(0))} 
\ln\left (\frac{Q^2}{\Lambda^2}
\right ) \ ,
\label{rQinfty}
\end{equation}
where the notation $\kappa_{T}(0)$ indicates that $\kappa_T$ was determined at $Q^2=0$. A comparison of (\ref{rQ0})
and (\ref{rQinfty}) suggests the interpolation formula
\begin{equation}
r_{T, {\rm intp.}}(Q^2, \kappa_T(0))=\frac{4\kappa_T(0)}{3(1-2\kappa_T(0))} \ln \left (c_1 \frac{Q^2}{\Lambda^2}+\exp(c_2)
\right ) \ ,
\label{ransatz}
\end{equation}
where 
\begin{equation}
c_2=\frac{3(1-2\kappa_{T}(0))}{4\kappa_{T}(0)} \ , 
\label{c_2}
\end{equation}
guarantees $r_{T, {\rm intp.}}(Q^2=0)=1$, while $c_1$ has to be adjusted by using the numerical 
integration of (\ref{sigTA}). We note that a value of $c_1 \approx 1.50$ will 
be obtained in the numerical analysis of \mbox{Section \ref{section5}}.

In order to proceed to the second step, let us suppose that an appropriate 
$Q^2$ dependence of $\kappa_T(Q^2)={\bar z}_T(Q^2)(1-{\bar z}_T(Q^2))$ 
inserted into (\ref{sigTA2}) will result in $r_{T}(Q^2, \kappa_T(Q^2))=1$ in the full range of
$Q^2$ from $Q^2=0$ to $Q^2 \to \infty$. Going again through the
arguments leading to the interpolation formula (\ref{ransatz}), one finds that
the functional form of $\kappa_T(Q^2)$ is found by requiring 
\begin{equation}
r_{T, {\rm intp.}}(Q^2, \kappa_T(Q^2))=\frac{4\kappa_T(Q^2)}{3(1-2\kappa_T(Q^2))} 
\ln\left (c_1\frac{Q^2}{\Lambda^2}+\exp(c_2)
\right )=1 \ .
\label{rQinftyQ}
\end{equation}
In fact, asymptotically, the expression for $r_{T, {\rm intp.}}(Q^2,\kappa_T(Q^2))$ in 
(\ref{rQinftyQ}) again
coincides with the ratio of (\ref{sigTA3}) and (\ref{sigTqaq}). Moreover, 
(\ref{rQinftyQ}) for $Q^2=0$ yields relation (\ref{c_2}) as the correct
constraint on $\kappa_T(Q^2)$ for $Q^2=0$. Solving (\ref{rQinftyQ}) for $\kappa_T(Q^2)$, we obtain
\begin{equation}
\kappa_T(Q^2)=\frac{3}
{6+4 \ln \left (c_1 \frac{Q^2}{\Lambda^2}+\exp(c_2) \right )} \ .
\label{kappaQ}
\end{equation}
In Section \ref{section5}, it will be explicitly shown that 
${\sigma}_{{\gamma}^{\ast}_T p }(W^2, Q^2; {\bar z}_T, \delta_T)$ from 
(\ref{sigTA2}), upon substituting the $Q^2$ dependence for $\kappa_T(Q^2)$ from (\ref{kappaQ}), will indeed provide an excellent
representation of the exact result calculated by numerical evaluation of
(\ref{sigTA}).  

If, instead of the $\delta$--function, the Gaussian (\ref{gauss}) is inserted for
${\tilde \sigma}_{(q{\bar q})p}$ in (\ref{sigTA}), the same averaging
procedure in the integrand leads to 
\begin{eqnarray}
{\sigma}_{ {\gamma}^{\ast}_T p }&&(W^2, Q^2; {\bar z}_T, \delta_T) 
 =  \frac{\alpha}{2 \pi} \left( \frac{e_q}{e_0} \right)^2 
{\sigma}^{(\infty)}_{(q{\bar q})p}R_0^2 (1-2{\bar z}_T(1\!-\!{\bar z}_T)) 
\int_{0}^{\infty} d {l}^2_{\perp} e^{-{l}_{\perp}^2 \cdot 
R_0^2} 
\times
\nonumber\\
&&
{\Bigg \lbrack} 
\left (
(1\!+\!2 \delta_T) 
\frac{Q^2}{{\bar l}^{\prime 2}_{\perp}}+(1\!+\!\delta_T)\right
) \ln \left (1+\frac{{\bar l}_{\perp}^{\prime 2}}
{(1\!+\!2 \delta_T)(Q^2+M_0^2({\bar z}_T))}\right )
- \frac{Q^2}{(Q^2+M_0^2({\bar z}_T))} 
{\Bigg \rbrack} \ .
\label{sigT2}
\end{eqnarray}
At this stage, it is legitimate to expand the second expression in the brackets
in powers of ${\bar l}_{\perp}^{\prime 2}/(Q^2\!+\!M^2_0({\bar z}_T))$, 
at least for large $Q^2$, because the ${\bar l}_{\perp}^{\prime 2}$ values are
suppressed due to the Gaussian function in the integrand. Doing this, we get in
the limit $Q^2 \to \infty$
\begin{equation}
{\sigma}_{ {\gamma}^{\ast}_T p }(W^2, Q^2 \to \infty; {\bar z}_T) 
= \frac{\alpha}{3 \pi} \left( \frac{e_q}{e_0}
\right)^2 {\sigma}^{(\infty)}_{(q{\bar q})p}\frac{3(1-2{\bar z}_T(1-{\bar
    z}_T))}{4 {\bar z}_T(1-{\bar z}_T)}
\left \lbrack \frac{1}{R_0^2 Q^2} + {\cal O}\left
 (\frac{1}{Q^4}\right )
\right \rbrack \ . 
\label{sigT3}
\end{equation}
This expression coincides with  (\ref{sigTA3}), if $\Lambda^2$ is identified
with ${\Lambda}^2=1/R_0^2$. The conclusion on the relevance of the
configuration variable $z$ for the true asymptotic $(\ln Q^2)/Q^2$ behaviour 
(\ref{sigTqaq}) of the cross section is independent of whether we choose
a Gaussian, or a $\delta$--function, or any other physically
reasonable function for the $(q{\bar q})p$ interaction function
${\tilde \sigma}_{(q{\bar q}p}(l^2_{\perp})$ appearing in (\ref{sigTA}).

\subsection{The longitudinal cross section, ${\sigma}_{{\gamma}^{\ast}_L p}$.}

As in the transverse case, the integration of the $\Theta$--independent
part of (\ref{sigLA}) over $dM^{\prime 2}$ can be carried out analytically. 
We then obtain
\begin{eqnarray}
&&{\sigma}_{ {\gamma}^{\ast}_L p }(W^2, Q^2) 
 = 2 \alpha 
\left( \frac{e_q}{e_0} \right)^2 Q^2
\int_0^{\infty} d l^2_{\perp} {\tilde \sigma}_{(q{\bar q})p}(l^2,W^2)
\int_0^1 d z z (1\!-\!z)
\times 
\nonumber\\
&&
{\Bigg \lbrace}\int_{M_0^2(z)}^{\infty} d M^2 \left \lbrack
\frac{1}{(Q^2+M^2)^2} -
\frac{1}{(Q^2+M^2)\sqrt{(Q^2+{M}^{\prime 2}+{\Lambda}^{\prime 2})^2-4
 {\Lambda}^{\prime 2} M^2}} \right \rbrack
\nonumber\\
&&
+\!\int_{M_0^2(z)}^{\infty} d M^2 \Theta(M_0^2(z)\!-\!(M\!-\!{\Lambda}^{\prime})^2)\frac{1}{\pi}
\int_{(M-{\Lambda}^{\prime})^2}^{M_0^2(z)} d M^{\prime 2} 
w(M^{2},M^{\prime 2}, {\Lambda}^{\prime 2}) 
\frac{1}{(Q^2\!+\!M^2)(Q^2\!+\!{M}^{\prime 2})}
{\Bigg \rbrace} \ . \nonumber\\
\label{sigLA4}
\end{eqnarray} 
The presence of the $\Theta$--function term in (\ref{sigLA4}), which behaves as
$1/Q^2$ for $Q^2\!\gg\!\Lambda^2$, just as the main term, does not allow one
to carry out a further step analytically.

Employing the mean-value theorem with respect to the integrations over $dz$ and
$dM^{\prime 2}$, inserting (\ref{intm1}) with $\delta_T$ replaced by 
$\delta_L$ [cf.~(\ref{Mmitt1})], and dropping 
the $\Theta$--function term, we get
\begin{eqnarray}
{\sigma}_{ {\gamma}^{\ast}_L p }(W^2, Q^2; {\bar z}_L, \delta_L) 
&=& 2 \alpha \left( \frac{e_q}{e_0} \right)^2 Q^2 {\bar z}_L (1\!-\!{\bar z}_L)
\int_0^{\infty} d l^2_{\perp} {\tilde \sigma}_{(q{\bar q})p}(l^2,W^2)
\int_{M_0^2({\bar z}_L)}^{\infty} d M^2 \times
\nonumber\\
&& 
\left \lbrack \frac{1}{(Q^2+M^2)^2} -
\frac{1}{ (Q^2+M^2) (Q^2+M^2+ {\bar l}_{\perp}^{\prime 2}/(1\!+\!2 {\delta}_L)) }
\right \rbrack \ .
\label{sigLA1}
\end{eqnarray}
Inserting the $\delta$--function ansatz (\ref{delta})
for ${\tilde \sigma}_{(q{\bar q})p}(l^2_{\perp})$ and carrying out 
the trivial integration over $d l^2_{\perp}$,
we find agreement with the destructive-interference ansatz of
off-diagonal GVD. Upon integration over $dM^2$, we find
\begin{eqnarray}
\lefteqn{
{\sigma}_{ {\gamma}^{\ast}_L p }(W^2, Q^2; {\bar z}_L, \delta_L) 
=\frac{2 \alpha}{\pi} 
\left( \frac{e_q}{e_0}\right)^2 {\sigma}^{(\infty)}_{(q{\bar q})p} Q^2 
{\bar z}_L (1\!-\!{\bar z}_L)
{\Bigg \lbrack} \frac{1}{(Q^2+M_0^2({\bar z}_L))}  
} \nonumber\\
&&
- \frac{ (1\!+\!2 \delta_L)}{ {\bar \Lambda}^{\prime 2} }
\ln \left( 1+\frac{ {\bar \Lambda}^{\prime 2} }
{ (1\!+\!2 \delta_L)(Q^2+M_0^2({\bar z}_L)) } \right )
{\Bigg \rbrack} \ .
\label{sigLA2}
\end{eqnarray} 
Expansion of the logarithm yields for $Q^2 \to \infty$ a $1/Q^2$
behaviour
\begin{eqnarray}
{\sigma}_{ {\gamma}^{\ast}_L p }(W^2, Q^2 \to \infty; \delta_L) 
= \frac{2 \alpha}{{\pi}} 
\left( \frac{e_q}{e_0} \right)^2 {\sigma}^{(\infty)}_{(q{\bar q})p} 
{\Bigg \lbrack} 
\frac{{\Lambda}^2}{2 (1\!+\!2 \delta_L)Q^2}
+ {\cal O}\left(\frac{1}{Q^4} \right)
{\Bigg \rbrack} \ .
\label{sigLA3}
\end{eqnarray}
In the $Q^2 \to 0$ limit we obtain the expected linear $Q^2$ dependence
\begin{eqnarray}
{\sigma}_{ {\gamma}^{\ast}_L p }(W^2, Q^2 \to 0; {\bar z}_L, \delta_L)  =
 \frac{2 \alpha}{{\pi}}&& \left( \frac{e_q}{e_0}
\right)^2 {\sigma}^{(\infty)}_{(q{\bar q})p} Q^2 {\bar z}_L(1\!-\!{\bar z}_L)
\times
\nonumber\\
&&
\left \lbrack \frac{1}{M_0^2({\bar z}_L)}-
\frac{ (1\!+\!2 \delta_L) }{ {\bar \Lambda}^{\prime 2} } 
\ln \left (1+\frac{{\Lambda}^{2}}{(1\!+\!2
      \delta_L) k_{\perp 0}^2} \right ) + {\cal O}(Q^2)\right \rbrack .
\label{sigLA2a}
\end{eqnarray}
In contrast to the transverse case, there is no analytical 
evaluation available, not even for $Q^2 \to \infty$. From the 
numerical integration to be
presented in Section \ref{section5}, we will see that, in contrast to the 
transverse case, (\ref{sigLA2}) practically coincides with the exact result,
even at $Q^2\!\gg\!\Lambda^2$. In other words, in distinction from the
transverse cross section, in the longitudinal case, the effective value, 
${\bar z}_L$, of the
configuration variable, $z$, turns out to be constant, independent of $Q^2$. 
The effective mean configuration of the $q{\bar q}$ system building up the cross section
is the same at all values of $Q^2$.

Combining (\ref{sigLA3}) with the analytical result (\ref{sigTqaq}) for 
${\sigma}_{ {\gamma}^{\ast}_T p }$, we obtain an asymptotic decrease of the 
longitudinal-to-transverse ratio $R \sim 1/\ln Q^2$
\begin{equation}
R \equiv \frac{{\sigma}_{{\gamma}^{\ast}_L p}}{{\sigma}_{{\gamma}^{\ast}_T p}}
 = \frac{3}{(1+2\delta_L) \ln \left( Q^2/\Lambda^2 \right ) } \ .
\label{Ratio}
\end{equation}

Evaluating the full expression (\ref{sigLA4}) 
numerically and equating the $Q^2 \to \infty$ result
with the GVD formula (\ref{sigLA3}) 
determines $\delta_L$. The slope of the $Q^2 \to 0$ limit
\begin{equation}
\left \lbrack\frac{d{\sigma}_{{\gamma}^{\ast}_L p }
(W^2, \frac{Q^2}{{\Lambda}^2}; {\bar z}_L, \delta_L)}{d \left (\frac{Q^2}{{\Lambda}^2}\right )}\right
\rbrack_{Q = 0}=\frac{2 \alpha}{\pi}\left (\frac{e_q}{e_0}\right )^2
{\sigma}^{(\infty)}_{(q{\bar q})p}\kappa_L \left \lbrack \frac{{\Lambda}^2
  \kappa_L}{k_{\perp 0}^2}
-(1+2\delta)\kappa_L \ln \left (1+
\frac{ {\Lambda}^2/k_{\perp 0}^2 }{(1+2 \delta_L)} \right ) \right \rbrack, 
\label{siqLA6}
\end{equation}
then determines 
$\kappa_L={\bar z}_L(1-{\bar z}_L)$. The numerical values are given in Table 
\ref{table}. 
As shown in Section \ref{section5}, the
mean-value evaluation (\ref{sigLA2}), with $Q^2$--independent values for 
$\kappa_L$ and $\delta_L$, 
practically agrees with the
exact evaluation. 

As in the transverse case, we may evaluate (\ref{sigLA1}) for the case of the
Gaussian ${\tilde \sigma}_{(q{\bar q})p}$ (\ref{gauss}). 
The asymptotic result coincides with (\ref{sigLA3}), provided the
identification $\Lambda^2=1/R_0^2$ is made.

\section{Numerical Evaluation of 
${\sigma}_{{\gamma}_{T,L}^{\ast}p}(W^2, Q^2)$.}
\label{section5}

An analytic procedure to carry out the four-fold integration in the
expressions (\ref{sigTA}) and (\ref{sigLA}) for ${\sigma}_{{\gamma}^{\ast}_Tp}$
and ${\sigma}_{{\gamma}^{\ast}_Lp}$ for arbitrary values of $Q^2$ is not
available. We will accordingly integrate (\ref{sigTA}) and (\ref{sigLA})
numerically and determine the mean values of 
the configuration variables, ${\bar z}_{T,L}$, and of the $M^{\prime}$--mass
variables $\delta_{T,L}$
by comparison of the numerical results with the mean-value evaluation. As mentioned, the full
expressions (\ref{sigTA}) and (\ref{sigLA}), including the low-mass
$\Theta$--function corrections are numerically integrated, the effect of the
$\Theta$--function term thus being absorbed in the numerical values of ${\bar
  z}_{T,L}$ and $\delta_{T,L}$.

For the numerical evaluation of (\ref{sigTA}) and (\ref{sigLA}), we again
specialize to the $\delta$--function ansatz (\ref{delta}) for 
${\tilde \sigma}_{(q{\bar q})p}$. The expression for 
${\sigma}_{{\gamma}^{\ast}_{T,L}p}$ in (\ref{sigTA}) and (\ref{sigLA}) may then
be rewritten in terms of the ratios of $Q^2/\Lambda^2$ and
$\Lambda^2/k_{\perp 0}^2$ and integrated numerically\footnote{
Actually, for the main 
term, the $dM^{\prime 2}$ integrations were carried out analytically, and the
$dz$, $dM^2$ integrations numerically, while for the $\Theta$--function term 
the three-fold integration over $dM^{\prime 2}$, $dz$, and $dM^2$ was done 
numerically.}. 

In the transverse cross section, ${\bar z}_T$ and $\delta_T$ are determined
by equating the numerical results for the cross section and its derivative with
respect to $Q^2$ at $Q^2 \approx 0$ with the mean-value formula (\ref{sigTA2}).

For the longitudinal cross section, the derivative with respect to $Q^2$ at 
$Q^2 \approx 0$ and the cross section for asymptotic values of $Q^2/\Lambda^2$ 
are used. The results of the analysis are presented in Table \ref{table}.

\begin{table}
\begin{center}
\begin{tabular}{|c|c|c|c|c|}\hline
${\Lambda}^2/k_{\perp 0}^2$ & $\delta_L$ & $\delta_T$ & ${\bar z}_L(1-{\bar
  z}_L)$ & ${\bar z}_T(1-{\bar z}_T)$ \\ \hline
4                         & 0.1009  & 0.4321  & 0.1624  & 0.2186 \\ \hline
2                         & -0.0651 & 0.3876  & 0.1686  & 0.2047 \\ \hline
1                         & -0.1767 & 0.5224  & 0.1714  & 0.1455 \\ \hline
\end{tabular} 
\end{center}
\caption{The parameters $\delta_{L,T}$ and $\kappa_{L,T}\equiv
{\bar z}_{L,T}(1\!-\!{\bar z}_{L,T})$ determined, as explained in text, for 
various ${\Lambda}^2/k_{\perp 0}^2$.}
\label{table}
\end{table}      

Turning to a discussion of the $Q^2$ dependence, we fix $\Lambda$ to the value
of $\Lambda^2\!=\!0.05 \ {\rm GeV}^2$. This value is suggested from  
$\Lambda^2\!=\!1/R_0^2$, if $R_0$ is identified with the proton radius, 
$R_0 \sim 1$ fm. A similar value for $\Lambda^2$ follows from the
identification of $\Lambda^2=l^2_{\perp}$ [cf.~(\ref{delta})], 
where the momentum transfer $l^2_{\perp}$ 
is transmitted by gluon exchange of order
$\Lambda^2_{\rm QCD}\!\sim\!0.01$-$0.1 \ {\rm GeV}^2$. 
We will usually use the same value for
the transverse extension 
of the incoming low-mass $q{\bar q}$ state, i.e. 
$\Lambda^2/k^2_{\perp 0}\!=\!1$, or
$k^2_{\perp 0}\!=\!0.05 \ {\rm GeV}^2$.

In Fig.~\ref{TVTzlog3100}, we show the ratio, as defined by (\ref{gerh}), of the result of the numerical
integration and the mean-value evaluation of the transverse cross section for
$\Lambda^2/k^2_{\perp 0}\!=\!1$ ($\Lambda^2\!=\!0.05 \ {\rm GeV}^2$) as a 
function of $Q^2$. As a consequence of determining 
$\kappa_T \equiv {\bar z}_T(1-{\bar z}_T)$ and $\delta_T$ at $Q^2=0$, the ratio
$r_T(Q^2)$ from (\ref{gerh}) equals unity at low $Q^2$, while showing
the logarithmic growth expected according to (\ref{rQinfty}) for $Q^2 \to
\infty$. As shown in Fig.~\ref{TVTzlog3100}, the interpolation formula
(\ref{ransatz}) with
\begin{equation}
\kappa_T(0)=0.1455, \quad c_1=1.50,\quad c_2=3.65
\label{kc1c2}
\end{equation}
yields an excellent representation of the functional form of the ratio.
Here, $\kappa_T(0)$ is given in Table \ref{table}, $c_2$ is obtained from
(\ref{c_2}), and $c_1$ was determined by requiring agreement of expression
(\ref{ransatz}) with the actual ratio (\ref{gerh}) at $Q^2 \gg \Lambda^2$.

In Fig.~\ref{TVTzlog3100}, we also show the ratio $r_T(Q^2, \kappa_T(Q^2))$ that is calculated by
inserting the $Q^2$ dependence from (\ref{kappaQ}) for the effective
value $\kappa_T(Q^2) \equiv {\bar z}_T(Q^2)(1-{\bar z}_T(Q^2))$ into the mean-value 
evaluation (\ref{sigTA2}). The (almost) constant value of $r_T(Q^2,
\kappa_T(Q^2))$ explicitly
shows that (\ref{sigTA2}), together with the effective $Q^2$ dependence of the
configuration variable, yields an excellent representation of the $Q^2$
dependence of ${\sigma}_{{\gamma}^{\ast}_Tp}$ from (\ref{sigTA}). 
The numerical results for $\kappa_T$ in Table \ref{table1}, obtained from
(\ref{kappaQ}), show how $\kappa_T(Q^2)$ and ${\bar z}_T(Q^2)$ decrease with
increasing $Q^2$. With increasing $Q^2$, a larger and larger part of the
transverse cross section is induced by $q{\bar q}$ configurations with small
angles in their rest frame relative to the virtual-photon direction. This shift
in the effective $q{\bar q}$ configuration is responsible for the logarithmic
scaling violation of the transverse part of the structure function $F_2$. 
\renewcommand{\textfloatsep}{1cm}
\begin{figure}[htb]
\setlength{\unitlength}{1.cm}
\begin{center}
\epsfig{file=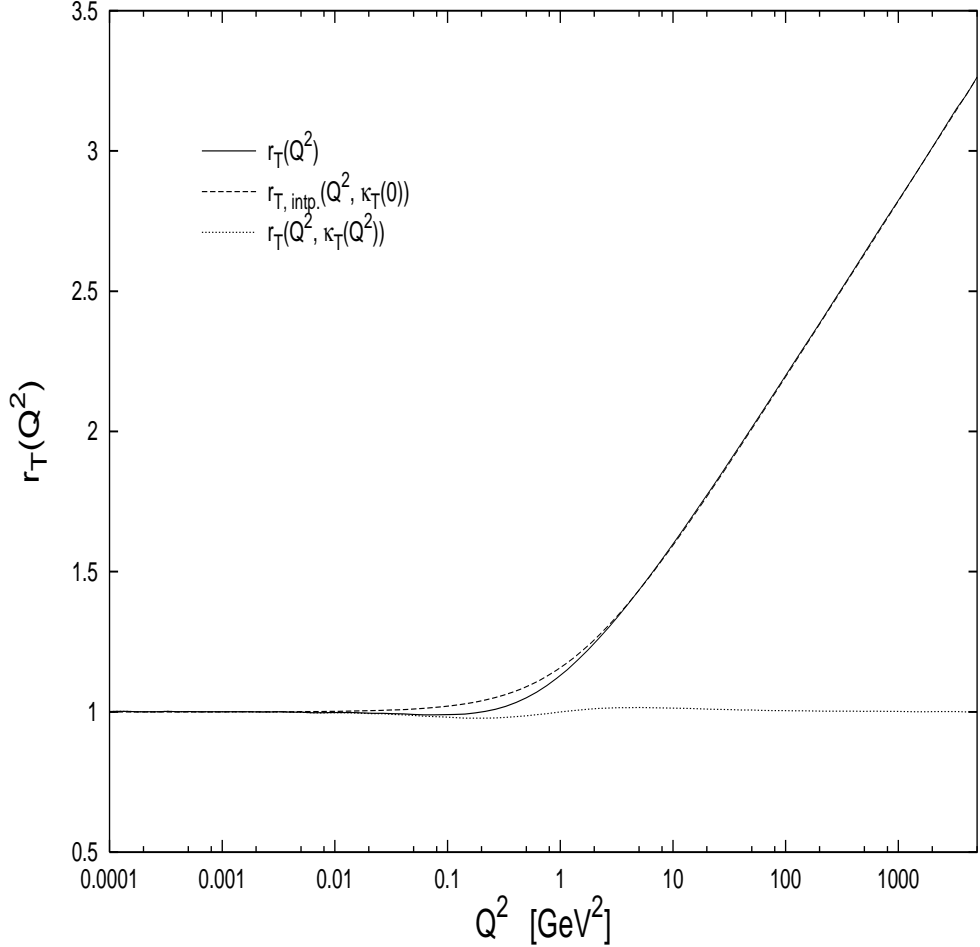, height=13cm, width=13cm}
\end{center}
\caption{\small The solid line shows the ratio 
$r_T(Q^2)={\sigma}_{{\gamma}^{\ast}_Tp}(W^2, Q^2)/
{\sigma}_{{\gamma}^{\ast}_Tp}(W^2, Q^2; {\bar z}_T, \delta_T)$ from 
(\ref{gerh}). The numerator is obtained by numerical integration of
(\ref{sigTA}), the denominator by evaluating the mean-value expression
(\ref{sigTA2}). The dashed line shows the result of the interpolation formula 
(\ref{ransatz}) with the parameters (\ref{kc1c2}). Finally, the dotted line
results from inserting $\kappa_T(Q^2)$ into the mean-value evaluation in the
denominator of $r_T(Q^2)$.}
\label{TVTzlog3100}
\end{figure}
In the approximation of a constant $Q^2$--independent value of $\kappa_T$, the
logarithmic scaling violation is evidently lost, while scaling remains. As
emphasised before, it is the cancellation between diagonal and off-diagonal
contributions (in mass) to the forward Compton amplitude, related to the
two-gluon exchange structure, that is responsible for scaling, and not the
effective change of the $q{\bar q}$ configuration with $Q^2$.
\begin{table}
\begin{center}
\begin{tabular}{|c|c|c|}\hline
$Q^2 \ [{\rm GeV}^2]$ & $\kappa_T(Q^2)$ & ${\bar z}_T(Q^2)$ \\ \hline
0.0001    & 0.1455 & 0.1767, 0.8233 \\ \hline
0.001     & 0.1455 & 0.1767, 0.8233 \\ \hline
0.01      & 0.1453 & 0.1764, 0.8236 \\ \hline
0.1       & 0.1434 & 0.1735, 0.8265 \\ \hline
1         & 0.1309 & 0.1549, 0.8451 \\ \hline
10        & 0.1024 & 0.1158, 0.8842 \\ \hline
100       & 0.0788 & 0.0862, 0.9138 \\ \hline
1000      & 0.0635 & 0.0681, 0.9319 \\ \hline
\end{tabular} 
\end{center}
\caption{The parameter $\kappa_{T}(Q^2)$ from (\ref{kappaQ}) and the related 
configuration variable ${\bar z}_{T}(Q^2)$ as a function of $Q^2$. We used ${\Lambda^2}/k^2_{\perp 0}=1$, or $k^2_{\perp 0}=0.05 \ {\rm GeV}^2$.}
\label{table1}
\end{table}   

In Fig.~\ref{LVL3100}, we show the results for the ratio, 
\begin{equation}
r_L(Q^2)\equiv
\frac{{\sigma}_{{\gamma}_{L}^{\ast}p(W^2, Q^2)}}
{{\sigma}_{{\gamma}_{L}^{\ast}p(W^2, Q^2; {\bar z}_L, \delta_L)}} \ ,
\label{r_L}
\end{equation}
of the numerical evaluation (\ref{sigLA}) and the
mean-value result (\ref{sigLA2a}) for the longitudinal cross section. This ratio is approximately equal to unity
over the whole range of $Q^2$; deviations from unity are of the order of
magnitude of $10\%$ for small values of 
$0.1 \ {\rm GeV}^2 \le Q^2 \le 10 \ {\rm GeV}^2$.
In the longitudinal case, the effective value of
$\kappa_L \equiv {\bar z}_L(1-{\bar z}_L)$ is independent of $Q^2$. In
contrast to the transverse case, it is the same $q{\bar q}$ configuration, 
with $\kappa_L = 0.1714$ for $\Lambda^2/k^2_{\perp 0}=1$ that determines the
cross section for arbitrary values of $Q^2$. The asymptotic scaling of the
longitudinal cross section together with the constancy of $\kappa_L$ 
explicitly shows that scaling is not related to an effective change in 
$\kappa_L$ with $Q^2$.
\renewcommand{\textfloatsep}{1cm}
\begin{figure}[htb]
\setlength{\unitlength}{1.cm}
\begin{center}
\epsfig{file=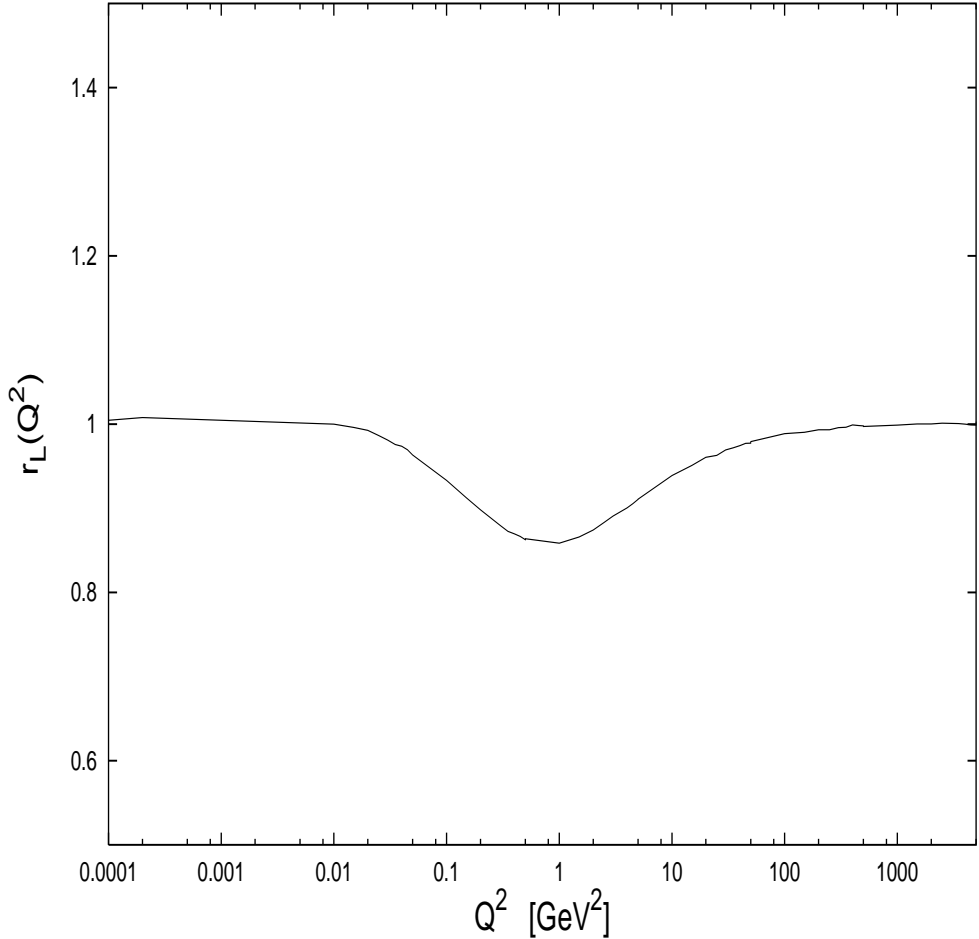, height=13cm, width=13cm}
\end{center}
\caption{\small The ratio $r_L(Q^2)={\sigma}_{{\gamma}^{\ast}_Lp}(W^2, Q^2)/
{\sigma}_{{\gamma}^{\ast}_Tp}(W^2, Q^2; {\bar z}_L, \delta_L)$ from
(\ref{r_L}). The numerator is the result of integrating (\ref{sigLA})
numerically, while the denominator is obtained from the mean-value result (\ref{sigLA2}).}
\label{LVL3100}
\end{figure}

In Fig.~\ref{TL3100num}, we show the numerical results for the transverse and
longitudinal cross sections normalised by (transverse) 
photoproduction as a function of $Q^2$. The results shown are obtained for 
$\Lambda^2=k_{\perp 0}^2=0.05 \ {\rm GeV}^2$. 

In view of the results in Figs.~\ref{TVTzlog3100} and \ref{LVL3100}, the numerical
integration of (\ref{sigTA}) and (\ref{sigLA}) and the mean-value evaluations 
(\ref{sigTA2}), with $\kappa_T(Q^2)$ from (\ref{kappaQ}), and (\ref{sigLA2}), respectively, practically
agree with each other. It is worth noting that the drop of the transverse cross
section by two orders of magnitude from $Q^2 \approx 0 \ {\rm GeV}^2$ to $Q^2
\approx 100 \ {\rm GeV}^2$ is of 
the order of magnitude seen in the experimental data \cite{HERA,SchildknechtSpiesberger}. It is not the aim of the
present paper to enter an analysis of the experimental data. Such an analysis
would require an extension of the present work by carefully incorporating the
$W^2$ dependence which is beyond the scope of the present work --
cf.~Refs.\cite{Shaw,Wuesthof}, and  
\renewcommand{\textfloatsep}{1cm}
\begin{figure}[htb]
\setlength{\unitlength}{1.cm}
\begin{center}
\epsfig{file=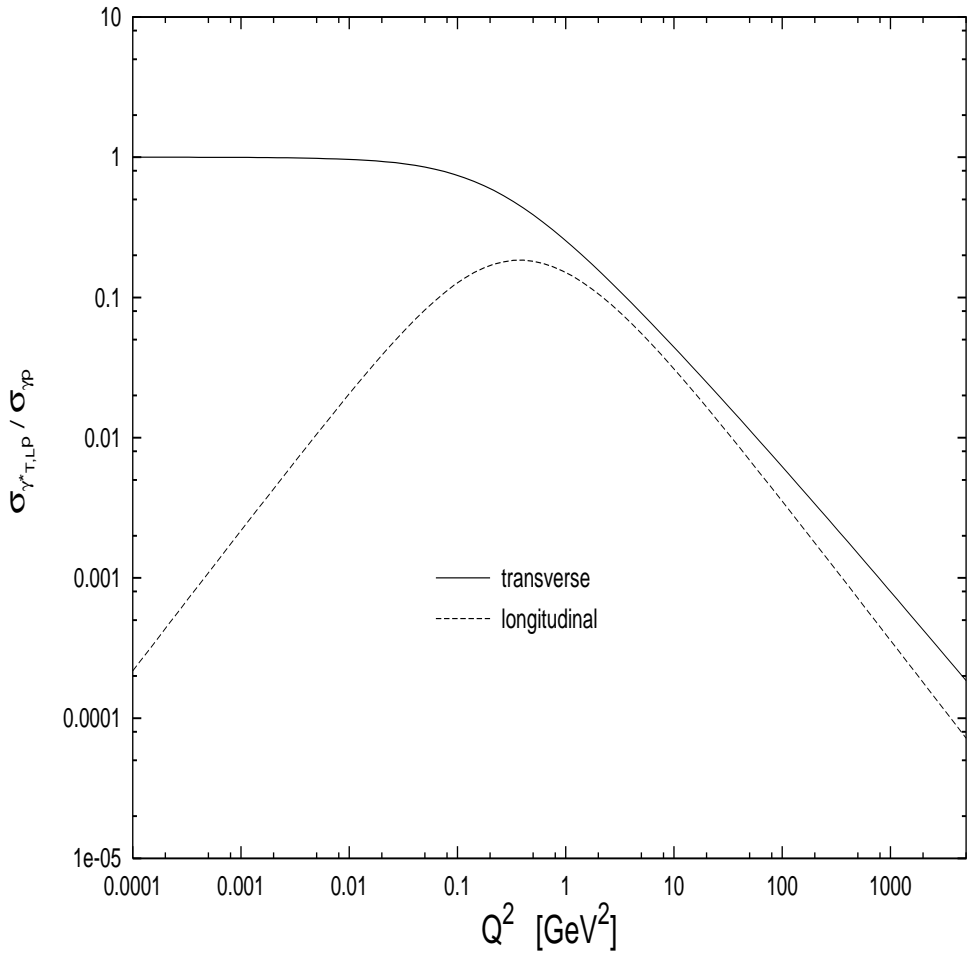, height=13cm, width=13cm}
\end{center}
\caption{\small Numerical results for ${\sigma}_{ {\gamma}^{\ast}_Tp }$
from (\ref{sigTA}) (solid line), and ${\sigma}_{{\gamma}^{\ast}_Lp }$ from (\ref{sigLA}) (dotted line), normalised by the
photoproduction cross section ${\sigma}_{{\gamma}p}$. The results shown are
obtained by numerical integration of (\ref{sigTA}) and (\ref{sigLA}). The
mean-value results from (\ref{sigTA2}) (with $\kappa_T(Q^2)$ from (\ref{kappaQ})) and (\ref{sigLA2}), respectively, coincide with the ones shown, apart from a minor
deviation in the longitudinal cross section around $Q^2 \approx 1 \ {\rm
  GeV}^2$ (compare Figs.~\ref{TVTzlog3100}, \ref{LVL3100}).} 
\label{TL3100num}
\end{figure}
Refs.~\cite{Gotsmanetal}-\cite{Bugaev}.

In Fig.~\ref{TLV3100}, we show the longitudinal-to-transverse ratio, 
\begin{equation}
R=\frac{{\sigma}_{{\gamma}_{L}^{\ast}p}}{{\sigma}_{{\gamma}_{T}^{\ast}p}} \ ,
\label{LVT}
\end{equation} 
as a function of $Q^2$. The solid curve shows the ratio of the cross sections 
from Fig.~\ref{TVTzlog3100}. The additional (dotted) curve shows the
effect of changing the threshold value of $k^2_{\perp 0}$. It was checked that
the change of $R$ with changing threshold is almost completely dominated by 
\renewcommand{\textfloatsep}{1cm}
\begin{figure}[htb]
\setlength{\unitlength}{1.cm}
\begin{center}
\epsfig{file=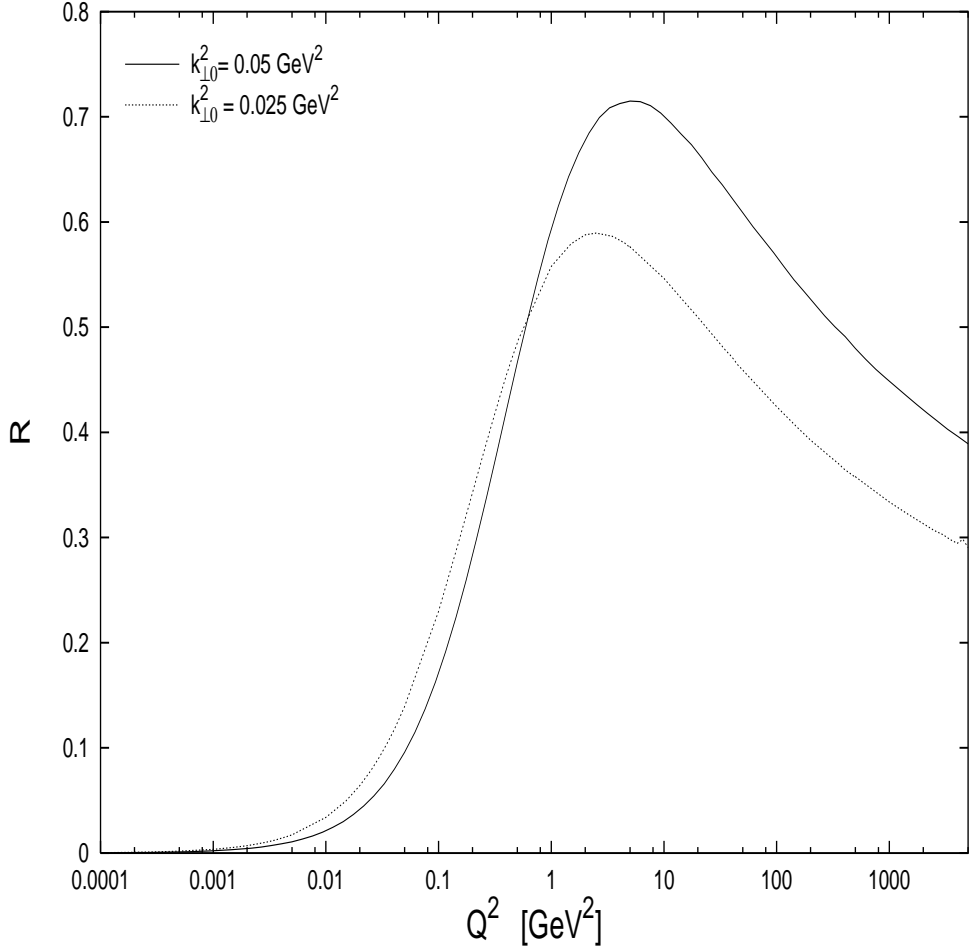, height=13cm, width=13cm}
\end{center}
\caption{\small The longitudinal-to-transverse ratio
  $R$ from (\ref{LVT}). The solid curve corresponds to 
${\Lambda^2}=k_{\perp 0}^2=0.05 \ {\rm GeV}^2$ as used in
Fig.~\ref{TL3100num}. The dotted curve is obtained for $\Lambda^2=0.05 \ {\rm
  GeV}^2$ and $k_{\perp 0}^2=0.025 \ {\rm GeV}^2$, as indicated.}
\label{TLV3100}
\end{figure}
the longitudinal cross section that decreases with decreasing threshold value.

We have also examined the effects on the results for 
${\sigma}_{{\gamma}^{\ast}_{T}p}$ and ${\sigma}_{{\gamma}^{\ast}_{L}p}$ induced
by the $z$ dependence of the threshold mass $M_0^2(z)$ from (\ref{M0z}) in 
(\ref{sigTA}) and (\ref{sigLA}). For this purpose we formed the ratio of
an evaluation with $z$--dependent threshold, $M_0^2(z)$, and an evaluation with
constant $M_0^2$. The latter threshold was chosen in such a way as to yield
a ratio equal to unity for $Q^2 \to 0$. While for $Q^2 \ge 100 \ {\rm GeV}^2$ 
the differences are well below $10\%$, they can reach values up to about 
$30\%$ for $Q^2 \approx 1 \ {\rm GeV}^2$ and up to $20\%$ at 
$Q^2 \approx 10 \ {\rm GeV}^2$. These effects have to be carefully considered 
in a comparison with the experimental data.
 
\section{Conclusions}
\label{section6}
We have provided a novel formulation of GVD for the low--$x$ diffraction region of 
deep inelastic scattering. The present work extends the GVD picture in so far as
the dependence on the internal structure of the ${\gamma}^{\ast} \to q{\bar q}$
transition is taken into account, and the ansatz for the scattering amplitude
for the strong $(q{\bar q})p$ interaction is inspired by the general structure
of two-gluon exchange. This ansatz implies a structure of destructive
interference in the forward Compton amplitude that was anticipated in off-diagonal GVD a
long time ago, and in fact, the present work provides a QCD-based a posteriori justification for
that ansatz. We have shown that the momentum-space formulation is identical to
a position-space formulation based on the concepts of a dipole cross section,
colour transparency and saturation.

The resulting $Q^2$ dependence has been cast into a fairly compact analytic 
form for arbitrary values of $Q^2$, including $Q^2=0$, by introducing effective
mean values for the configuration of the $q{\bar q}$ system, ${\bar z}$, and
also (as far as off-diagonal transitions are concerned) for its mass. It turned
out that the exact $Q^2$ dependence of the longitudinal cross section is well
represented by a $Q^2$--independent configuration, ${\bar z}_L$. In contrast, 
in the case of the transverse cross section, the effective mean value, 
${\bar z}_T$, of the $q{\bar q}$ configuration changes logarithmically with
$Q^2$. This logarithmic change of the effective configuration is responsible for a logarithmic violation of scaling of the 
structure function $F_2$.

In GVD, the $Q^2$ dependence of deep inelastic scattering is associated with the
propagation of (hadronic) $q{\bar q}$ states. While this principal feature of
GVD is retained, taking into account the structure of the $q{\bar q}$ system
explicitly, and using a QCD-inspired ansatz for $(q{\bar q})p$ scattering, 
leads to a logarithmic modification of the $1/Q^2$ dependence of the transverse
cross section of the original formulation of off-diagonal GVD. 
Asymptotically we have 
${\sigma}_{{\gamma}_T^{\ast}p}\!\sim\!(\ln Q^2)/Q^2$ corresponding to a
logarithmic violation of scaling for the structure function $F_2$. Moreover, the
longitudinal-to-transverse ratio,
$R\!\equiv\!{\sigma}_{{\gamma}_L^{\ast}p} /{\sigma}_{{\gamma}_T^{\ast}p}$,
decreases asymptotically as $1/\ln Q^2$.

\end{document}